\documentclass[twocolumn]{aastex62}

\usepackage{hyperref}
\usepackage{amsmath}

\received{January 1, 2018}
\revised{January 7, 2018}
\accepted{\today}
\submitjournal{ApJ}

\shorttitle{The History of Galaxy Minor Mergers}
\shortauthors{Conselice et al.}
\def\solm{M$_{\odot}\,$}

\def\solm{M$_{\odot}\,$}

\def\casgm20{CAS-G-M$_{20}\,$}
\def\m20{M$_{20}\,$}

\begin{document}

\title{A direct measurement of galaxy major and minor merger rates and stellar mass accretion histories at $z < 3$ using galaxy pairs in the REFINE survey}

\correspondingauthor{Christopher J. Conselice}
\email{conselice@manchester.ac.uk}

\author[0000-0002-0786-7307]{Christopher J. Conselice}

\affiliation{Jodrell Bank Centre for Astrophysics, Department of Physics and Astronomy, University of Manchester, Oxford Road, Manchester M13 9PL, UK}
\affiliation{Centre for Astronomy and Particle Physics, School of Physics and Astronomy, University of Nottingham, NG7 2RD, UK}
\author{Carl J. Mundy}
\affiliation{Centre for Astronomy and Particle Physics, School of Physics and Astronomy, University of Nottingham, NG7 2RD, UK}
\author{Leonardo Ferreira}
\affiliation{Centre for Astronomy and Particle Physics, School of Physics and Astronomy, University of Nottingham, NG7 2RD, UK}
\author[0000-0001-6889-8388]{Kenneth Duncan}
\affiliation{Centre for Astronomy and Particle Physics, School of Physics and Astronomy, University of Nottingham, NG7 2RD, UK}



\begin{abstract}

We measure the role of major and minor mergers in forming the stellar masses of galaxies over $0<z<3$ using a combination of $\sim 3.25$ deg$^{2}$ of the deepest ground based near-infrared imaging taken to date as part of the REFINE survey.  We measure the pair fraction and merger fractions for galaxy mergers of different mass ratios, and quantify the merger rate with newly measured time-scales derived from the Illustris simulation as a function of redshift and merger mass ratio.      We find that over $0 < z < 3$ major mergers with mass ratios greater than 1:4 occur $0.85^{+0.19}_{-0.20}$ times on average, while minor mergers down to ratios of 1:10 occur on average $1.43^{+0.5}_{-0.3}$ times per galaxy.  We also quantify the role of major and minor mergers in galaxy formation,  whereby the increase in mass due to major mergers is $93^{+49}_{-31}$\% while minor mergers account for an increase of $29^{+17}_{-12}$\%; thus major mergers add more stellar mass to galaxies than minor mergers over this epoch.  Overall, mergers will more than double the mass of massive galaxies over this epoch.   Finally, we compare our results to simulations, finding that minor mergers are over predicted in Illustris and in semi-analytical models, suggesting a mismatch between observations and theory in this fundamental aspect of galaxy assembly.  

\end{abstract}



\section{Introduction} \label{sec:intro}

The major contemporary idea concerning structure formation in the universe, including that of galaxies, is that they form hierarchically within a milieu of cold dark matter \cite[][]{blumenthal1984}. That is, the galaxies that we see today start from very small initial `seeds' of formation that later merge together to form larger and larger systems.  The evidence for this has however been lacking until mergers of galaxies with similar masses started to be examined and compared with theoretical models \citep[e.g.,][]{Carlberg1994,Patton1997,Conselice2003, Kitzbichler2008, Bertone2009, Jogee2009, Bluck2009, Mundy2017, duncan2019, Whitney2021,bickley2021, husko2022}.

A wealth of observations have also shown that the total stellar masses of galaxies increases by several factors over the past $\sim$10 Gyr \citep[e.g.,][]{Daddi2005, Ownsworth2016, Mortlock2015, leja2020, mcleod2021}. This is most obvious in the evolution of the number density of galaxies by a factor of $\sim10$ for the most massive galaxies at $z < 3$ \citep[]{Mortlock2015, conselice16}. A major further question is answering {\em how} these massive galaxies grown over time, that is what are the {\em physical processes} which add mass to these galaxies over the history of the universe? Effectively, what is the balance between star formation, mergers, and feedback?   There are two main pathways through which a galaxy increases its stellar mass: the process of star-formation and the consumption of other galaxies through merging.   It is therefore critical for understanding galaxy formation to disentangle the relative contribution from each formation method -- mergers and star formation -- in the build-up of stellar mass in massive galaxies.

For example, \cite{Mundy2015} demonstrate that a number density selection traces the same galaxy population better than a stellar mass selection.  Studies such as  \cite{Ownsworth2016} use number density selections at $z < 3$ to show that the star-formation history of massive galaxies can not fully account for the observed stellar mass growth at $z < 3$.  This can also be seen in the evolution of the galaxy stellar mass function, and the total number densities of galaxies which drops by a factor of $\sim 10$ from high redshift to low \citep[e.g.,][]{Mortlock2015, conselice16, mcleod2021}.  The role of mergers is thus crucial for galaxy formation, as without understanding these mergers, observed changes in galaxy masses and numbers cannot be accounted for. 

While many studies have investigated major mergers out to high redshift \citep[]{Conselice2003, Bluck2009,Bluck2012, Man2014, Mundy2015, Mantha2018, duncan2019,Ferreira2020,Dai2021, husko2022, shibuya2022}, the same cannot be said yet of mergers with smaller host-to-companion stellar mass ratios.  These are the the so-called {\em minor} mergers which are typically defined as galaxy pairings with $0.1 < \mu = (M1/M2) < 0.25$. The deep imaging, increased sample sizes, and high completeness needed to study these minor mergers has made it such that only recently has it been possible to study the evolution of minor mergers and their role in galaxy formation \citep{Bluck2012, Man2016A, Lofthouse2017, Shah2020}.  We currently have almost no observational constraints concerning this mode of galaxy formation for complete and statistically significant samples.  Furthermore, minor mergers are predicted in some cosmological models to be a significant method for forming galaxies, possibly triggering AGN and central black hole formation, and for instigating star formation \citep[e.g.,][]{Kaviraj2015, Shah2020}.  Furthermore, to fully understand galaxy formation, and test cosmological models in detail, the minor merger rate will need to be measured and fully accounted for.  

Minor mergers have also been increasingly implicated in the morphological and structural evolution of galaxies, including the observed size evolution of massive elliptical galaxies \citep[]{Trujillo2007, Buitrago2008, Bluck2009, Bluck2012, Ownsworth2016}. An apparent change in size of these galaxies by up to a factor of 5 from $z\sim3$ to $z\sim0$ can be theoretically explained by several processes. These include adiabatic expansion (`puffing up') from stellar mass loss or feedback mechanisms \citep[e.g.,][]{Fan2008,Damjanov2009}, or dry, dissipationless mergers. Simple arguments using the virial theorem suggest that the latter mechanism can produce an increase in size proportional to the square of the change in stellar mass \citep[e.g.,][]{Naab2009} from a merger event.     \cite{Bluck2012} using the GOODS NICMOS Survey (GNS) found that minor mergers can account for most of the observed size evolution at $z < 1$ if the merger timescale was sufficiently short at $\leq 1$ Gyr.  This suggests that minor mergers may be much more efficient at changing the size of a galaxy than mergers at larger stellar mass ratios \citep[e.g.,][]{Bezanson2009, Hopkins2010}.

Given a high enough rate of minor merger events, they may in fact be a dominant driver of the size and mass evolution of massive galaxies if a large fraction are so-called dry mergers without significant gas or star formation (e.g., \citep[][]{Bluck2012}). Indeed, we might expect that the fraction of galaxies undergoing a minor merger would be larger than that for major mergers, due  to the longer dynamical friction timescales between galaxies at these stellar mass regimes, and the fact that there are many more low mass galaxies than high mass galaxies. However, some cosmological simulations have indicated that major and minor merger rates are comparable (within a factor of $\sim2$) at high stellar masses ($>10^{11}\ \mathrm{M}_\odot$) and at redshifts of $z \leq 3$ \citep[]{Croton2006,Maller2006,Somerville2015}.

The simplest measurement that can be performed to investigate minor mergers is to measure the fraction of galaxies undergoing such events.  For example, \cite{Man2016A} used 3DHST/CANDELS and UltraVISTA observations to determine the role of minor mergers out to $z = 2.5$ for galaxies at $>10^{10.8}\ \mathrm{M}_\odot$. Ignoring a selection in flux rather than stellar mass, whereby the former generally selects gas-rich pairings that the latter would otherwise exclude, \cite{Man2016A} find that minor merger pair fractions are comparable to the major mergers, and exhibit a similar evolution with respect to redshift.   While \cite{Jogee2009} find that the fraction of morphologically selected minor mergers is at least 3 times that of major mergers out to $z\sim0.8$.  However, these results rely on the ability to visually distinguish mergers from star formation at low stellar masses.  

Probing galaxy mergers in a more indirect manner, \cite{Ownsworth2014, Ownsworth2016} find that minor mergers are the dominant source of stellar mass growth in the progenitors of modern $>10^{10.2}\ \mathrm{M}_\odot$ galaxies, responsible for forming $34\pm14 \%$ of the stellar mass during this time.   The results of these previous studies highlight the need for new analyses that combine wide-area, deep independent fields for the robust determination of the minor merger histories of galaxies.  We thus present results from an extensive new analysis of the total, major, and minor merger histories of massive galaxies at $0 < z < 3.5$.  We compare the major and minor merger histories, including their rates and the mass assembly due to various types of mergers.  We thus determine the role of various mergers in driving major aspects of galaxy formation and assembly.   We outline our methodology, results, and conclusions regarding the role of mergers in the formation of galaxies.  

The outline of this paper is as follows: \S \ref{sec:data} gives a description of the data and the methods in which we measure our stellar masses and photometric redshifts described in \S \ref{sec:uds}. \S \ref{sec:data_products} describes the various data products, while \S \ref{sec:results} contains our results, which includes a new measurement of the merger time-scale for galaxies and as a result the measurement of merger and stellar mass accretion rates.  In \S \ref{sec:discussion} we discuss our results in terms of the formation of galaxies as measured in an empirical way, including  the calculation of the stellar mass accreted through mergers,  Section \S \ref{sec:conclusion} summarises and concludes the results of this paper.

\section{Data and Fields}
\label{sec:data}

\subsection{Data Sets}

The work in this paper utilises several datasets from the REFINE survey (Redshift Evolution and Formation in Extragalactic Systems) \citep[][]{Mundy2017, Sarron2021} to achieve a measurement of the total and minor merger histories of massive galaxies.   This survey is described in detail in \cite{Mundy2017} (hereafter M17), however a brief description follows.

At low redshift, multi-wavelength photometry and spectroscopic observations from the second data release (DR2) of the GAMA survey are used. This full dataset provides a flux-limited sample of galaxies from three independent lines of sight (totalling 144 square degrees) down to a limiting Petrosian $r$-band magnitude of $m_r = 19$ \citep[][]{Driver2011}.

At higher redshifts $0.2 < z < 3$, three independent lines of sight provide an effective area of 3.25 square degrees - namely the UDS, VIDEO and Ultra-VISTA data sets. This includes the eighth data release (DR8) of the UKIDSS UDS provides galaxies down to a limiting $K$-band magnitude of $m_K = 24.6$ over 0.77 square degrees. The UKIDSS UDS remains the deepest $K$-selected square-degree sized survey to date. The VIDEO survey, combined with CFHT observations in the CFHT-LS D1 field, provides multi-wavelength observations over a $\sim$ 1 square degree field. Finally, UltraVISTA observations over 1.62 deg$^{2}$ in the COSMOS field are combined with archival observations. The publicly available catalogue presents a sample of galaxies down to a limiting magnitude of $m_K = 23.4$.   We describe these fields in more detail below.

From these datasets photometric redshift probability distributions, $P(z)$, are calculated using {\tt{EAZY}}, and stellar mass-redshift functions, $\mathcal{M}_*(z)$, are calculated using a custom spectral energy distribution (SED) fitting routine.  The full description of this data and how we reduced and analyzed the stellar masses and redshifts is described in M17. We provide a summary in \S \ref{sec:data_products} for how these are computed.

\subsection{UKIDSS Ultra Deep Survey (UDS)}
\label{sec:uds}

We use the eighth data release (DR8) of the UKIDSS UDS (e.g., \cite{Krishan2020}, Almaini et al. \textit{in prep}) in this paper.   The UDS is the deepest of the UKIRT (United Kingdom Infra-Red Telescope) Infra-Red Deep Sky Survey \citep[UKIDSS; ][]{Lawrence2007} project.  The UDS covers $0.77$ square degrees in total and includes photometry  in the the NIR $J$, $H$ and $K$  bands down to limiting $5\sigma$ AB magnitudes of 24.9, 24.2 and 24.6 as measured in 2'' apertures.  We utilize other multi-wavelength observations in the form of: $u$-band data from CFHT Megacam, as well as deep $B$, $V$, $R$, $i$ and $z$-band data from the Subaru-XMM Deep Survey \citep{Furusawa2008}, $Y$-band data originates from the ESO VISTA Survey Telescope; and IR photometry from the Spitzer Legacy Program.   From our tests of the data we conclude that there are uncertainties on $K$-band photometry of $f_\lambda / \delta f_\lambda \approx 500$ ($\approx 5$) at $K = 19$ ($24$).   The data we use in this field gives us a wavelength coverage from $0.3 \mathrm{\mu m} < \lambda < 4.6 \mathrm{\mu m}$. 

The $K$-band based catalog of this field contains approximately 90,000 galaxies out to $z \sim 3.5$, reaching a $99\%$ completeness depth of $K = 24.3$ within 0.63 square degrees. We furthermore use spectroscopic redshifts from archival sources, and the UDSz \citep{Curtis-Lake2012,Bradshaw2013}.  This gives us 2292 high quality spectroscopic redshifts at $0 < z < 4.5$ (90\% at $z < 2$) in this field alone. We use this spectroscopic redshift sample to calibrate our photometric redshift probability distributions, as well as our stellar mass measurements.

\subsection{UltraVISTA}
\label{sec:uvista}

We furthermore utilize the publicly available  UltraVISTA catalogue, which is also $K_s$-band selected.     The UltraVISTA is the location for the HST COSMOS field \citep{Scoville2007b}, and thus is of interest for studies involving the structures of galaxies.    The UltraVista field covers an effective area of $1.62$ deg$^{2}$, and we use a catalog with PSF-matched $2.1''$ aperture photometry across 30 bands covering the wavelength range $0.15 \mathrm{\mu m} < \lambda < 24 \mathrm{\mu m}$.  We calculate a $90\%$ completeness limit at $K_s = 23.4$. To obtain a clean sample we only use sources above this detection limit. We find that the typical uncertainties on the $K_s$-band photometry are $f_\lambda / \delta f_\lambda \approx 200$ ($\approx 10$) at $K_s = 19$ (23). 

We also derive photometric redshifts and stellar masses as described later in this paper, and in M17.   To carry this out we use matched catalogs from GALEX \citep{Martin2005}, CFHT/Subaru \citep{Capak2007}, S-COSMOS \citep{Sanders2007} and UltraVISTA \citep{McCracken2012}. We also use the  zCOSMOS Bright \citep{Lilly2007a} spectroscopy, which provides 5467 high quality spectroscopic redshifts at $z < 2.5$. The vast majority (99\%) of these spectroscopic redshifts are at $z < 1$ and 50\% are at $z < 0.5$.

\subsection{VIDEO}
\label{sec:video}

We furthermore use data from the VISTA Deep Extragalactic Observations (VIDEO) survey \citep{Jarvis2012}, which is a $\sim$12 square degree survey in the near-infrared $Z$, $Y$, $J$, $H$ and $K_s$ bands.  Like UltraVISTA the near-infrared data from VISTA is matched to data from the Canada-France-Hawaii Telescope Legacy Survey Deep-1 field (CFHTLS-D1). This provides multi-wavelength ($0.3\mathrm{\mu{m}} < \lambda < 2.1\mathrm{\mu{m}}$) coverage over a total of $\sim 1$ square degree.  The 90\% completeness is at magnitude of $K_s = 22.5$.   To determine this completeness we carry out a series of comprehensive simulations to calculate the detection completeness as a function of total $K$-band magnitude,  described in detail in M17.  We find that typical $K_s$-band uncertainties are $f_\lambda / \delta f_\lambda \approx 200$ ($\approx 15$) at magnitudes $K_s = 19$ (22).

For this field we use a $K_s$-selected catalog containing 54,373 sources after stellar removal using a $uJK$ colour selection, magnitude cuts, star masking, and retaining sources which have a detection signal-to-noise $> 2$. We mask out bright stars and contaminated areas manually using the VIDEO $K_s$-band image. The objects we locate within these masked regions are flagged and discarded from the sample. Within this field a spectroscopic sample of galaxies is constructed from the VIMOS VLT Deep Survey \citep{LeFevre2005}, and the VIMOS Public Extragalactic Redshift Survey \citep{Garilli2014} data releases. We only match our sample with the secure redshifts with high quality quality flags 3 and 4.  The criteria we use is to have matches within one arcsecond of our $K_s$-band sources. Overall, this provides 4,382 high-quality spectroscopic redshifts over the range $0 < z < 4.5$, with the majority (90\%) of systems below $z < 1.5$.  Further details on all three fields and our measure photometry and photometric redshfits are discussed in detail in M17.

\subsection{GAMA}
\label{sec:gama}

In order to obtain a measurement of the merger fraction at low redshifts the second data release (DR2) of the Galaxy And Mass Assembly (GAMA) campaign \citep{Driver2009,Liske2015} is used. This data provides multi-wavelength photometry in 9 filters over three fields totalling 144 square degrees. Complimenting this data, 98\% of the detections have secure spectroscopic redshifts. GAMA therefore represents a large and unique dataset with which to probe galaxy evolution at low redshift.

We use data from all three of the GAMA fields (G09, G12 and G15), which in this paper we call the GAMA region. When calculating stellar masses in this region, the recommended photometric zero-point offsets\footnote{\url{http://www.gama-survey.org/dr2/schema/table.php?id=168}} and stellar mass scaling factors \citep{Taylor2011} provided with the release documentation are applied. What differentiates this dataset from the others used in this work is the unprecedented spectroscopic coverage. Combining the three aforementioned GAMA regions yields 55,199 objects with good quality spectroscopic redshift (quality flag $\mathrm{nQ} > 2$ which provides spectroscopic redshifts at $>90\%$ confidence \citealt{Driver2011}) and $z_\mathrm{spec} > 0.005$, which minimises contamination from stars (visual inspection of a $(u-J)$ vs $(J-K)$ plot reveals this cut removes the stellar locus), representing 97 per cent of the total number of objects down to a limiting Petrosian $r$-band magnitude of $m_r = 19$. This allows the analysis to be performed in two ways: photometrically and spectroscopically, which give similar results within the calculated errors.  Typical uncertainties on $r$-band photometry are $f_\lambda / \delta f_\lambda \approx 700$ ($\approx 200$) at $r = 17$ (19).

\subsection{Simulated Data}
\label{sec:simdata}

Models of galaxy formation and evolution have advanced dramatically over the last few decades. Semi-analytic models (SAMs) aim to reproduce and predict the statistical properties of galaxy populations, historically at low redshift. In this work we use a few models to compare with our observations. This includes the Munich `family' of models \citep[e.g.,][]{Croton2006,DeLucia2006,Guo2011}, as described in \citet{Henriques2014}.  We use these models to compare theory predictions of the pair fraction with our observed ones. This model is applied to the output of The Millennium Simulation \citep{Springel2005}, scaled to a Planck cosmology \citep{Planck2014params}. We use all 24 mock lightcones from the German Astrophysical Virtual Observatory \citep[GAVO;][]{Lemson2006}, which are reduced in size from a circular aperture of two degrees diameter to a square field-of-view with an area of one square degree. Doing so permits one to quantify the expected variance between surveys similar in size to those used in this study. Finally, predictions of the merger rate and stellar mass accretion rate within the Illustris simulation \citep{Vogelsberger2014,Vogelsberger2014a,Rodriguez-Gomez2015} are compared to observational measurements.

\section{Data Products}
\label{sec:data_products}

In this section the photometric redshifts and stellar masses derived from the data sets described in Section \S \ref{sec:data} are explained.

\subsection{Photometric redshifts}
\label{subsec:photozs}

We calculate photometric redshift probability distributions for all sources using the {\tt{EAZY}} photometric redshift code \citep{Brammer2008a}.   The {\tt{EAZY}} code determines the $z_\text{phot}$ for each galaxy by fitting spectral energy distributions (SEDs) produced by a linear combination of templates to a set of photometric measurements. We use the default set of six templates, derived from the PEGASE models \citep{Fioc1999}, in combination with an additional red template from the \citet{Maraston2005} models, and a combination of single-burst \citet{Bruzual2003} templates of different ages.  This is required to provide robust SED fits to the diversity of observed galaxies in modern surveys \citep[e.g.,][]{Onodera2012, Muzzin2013a}.

We use this set of templates to calculate photometric redshifts and photometric redshift probability distributions (PDFs). The PDF is constructed for each galaxy from its $\chi^2(z)$ distribution following $P(z) \propto \exp(-\chi^2(z)/2)$, after convolution with a photometric prior. The following paragraphs discuss the use of a photometric prior in these calculations and the ability of the resulting PDFs to accurately reproduce photometric redshift confidence intervals.

In calculating galaxy PDFs and best-fit photometric redshifts, many studies make use of a luminosity or colour dependent redshift prior. The use of such priors have been shown to improve best-fit solutions when compared to spectroscopic redshift measurements \citep[e.g.,][]{Benitez2000,Brammer2008a}. However, the use of such priors may introduce biases into the measurement of close pairs. As an example, consider two galaxies at the same redshift with identical properties except for stellar mass (luminosity). A luminosity based prior will influence the probability distribution of each galaxy and, in this example, the higher mass system will have its PDF biased towards lower redshifts, and vice-versa for the second galaxy. Furthermore, priors are necessarily based on simulations. At higher redshifts ($z > 2$) these may deviate from the true distribution of galaxies, however at lower redshift they are much more constrained and in agreement with observations.

As such, a luminosity prior, $\text{P}(z|m)$, is constructed which denotes the probability of a galaxy with apparent $K$-band magnitude $m$ being found at redshift $z$, by extracting galaxy number counts from the \citep[][]{Henriques2014} semi-analytic model using 24 independent light cones. This model has been shown to accurately reproduce the observed number densities of galaxies out to $z \sim 3$, and thus is used to construct a prior from. This is achieved in the same manner as \citet{Brammer2008a} and \citet{Benitez2000}, parameterizing each magnitude bin $i$ as:

\begin{equation} \label{eqn:prior}
	P(z|m_{K,i}) \propto z^{\gamma_i} \times \exp( -(z/z_i)^{\gamma_i} ),
\end{equation}

\noindent where $\gamma_i$ and $z_i$ are fit to the redshift distribution in each magnitude bin. This ensures that the prior is smooth over the redshift range of interest. We calculate these distributions over the redshift range $0 < z < 7$ and an apparent magnitude range of $17 < m_K < 27$. 
Although we find that pair fractions obtained using photometric redshifts calculated with and without a prior are indistinguishable within the calculated uncertainties, we use this prior in this work as it improves the best-fit $z_\text{phot}$ estimates and reduces the number of catastrophic outliers (see M17).

\subsubsection{Photometric redshift confidence intervals}
\label{mergers:products:photoz:confidence}

Redshift probability distributions output by photometric redshift codes are often unable to accurately represent photometric redshift confidence intervals \citep[e.g.,][]{Hildebrandt2008,Dahlen2013}. The causes for this include inaccurate errors on the photometry or the limitations on the choice of template sets used in the fitting. Although average agreement between best-fit $z_\text{phot}$ and $z_\text{spec}$ can be excellent, 1$\sigma$ and 2$\sigma$ confidence intervals can be significantly over- or under-estimated.

Analysing the photometric redshift probability distributions output by {\tt{EAZY}}, discussed in Section \S \ref{subsec:photozs}, we find that the confidence intervals are incorrect. Using high quality spectroscopic redshifts for a subset of galaxies in each field. We find that 72\%, 71\%, 81\% and 50\% of $z_\text{spec}$ are found within the $1\sigma$ photometric PDF interval for the UDS, VIDEO, COSMOS and GAMA regions, respectively. In order to address this, PDFs that overestimate the confidence intervals are sharpened. The method we use is very similar to that carried out as in \citet{Dahlen2013}.

To carry out this sharpening the PDFs are replaced with a form $P(z_i) = P(z_i)_{0}^{1/\alpha}$ until the distribution gives the correct fraction of 68.3\%. To smooth, the PDFs are convolved with a kernel of $[0.25, 0.5, 0.25]$ until the correct fraction of 68.3\% is recovered. The same process is then applied to the entire sample. In doing so, values of $\alpha = 0.850$, $0.840$, $0.510$ are obtained for the UDS, VIDEO and COSMOS fields, respectively. However, the PDFs within the GAMA region are not smoothed as such a process leads to a high probability at the lowest redshifts which boosts the measured merger fraction in the lowest redshift bin. This is likely an artefact of using a linear redshift grid, however this is not expected to significantly affect the results of this work, as using unsharpened redshift PDFs in other fields results in no significant changes to measured pair fractions.

\subsubsection{Best-fit solutions}

While we mainly use the PDFs of stellar mass and redshift associated with each galaxy, we also compare best-fit photometric redshift solutions with spectroscopically obtained values.   The comparison between spectroscopic and photometric redshifts are shown in Figure~1.  To specify the quality of our redshifts in this work we use the normalised median absolute deviation (NMAD), mean $|\Delta{z}| / (1+z_\text{spec})$, where $\Delta{z} = (z_\text{spec} - z_\text{phot})$, and the outlier fraction, defined in two ways (see Table~1).  Various other ways to quantify the photometric redshift quality are discussed in detail in M17.   In summary, we find that  all fields, except for GAMA, possess averages biases of $(z_\text{spec} - z_\text{phot}) \approx 0$.   We also find a relatively large apparent bias in our photometric redshifts within the GAMA region whereby our photometric redshifts tend to be larger than the spectroscopic redshift by an average of $\Delta{z} = 0.02$.  This issue is however not expected to affect the results presented herein, as discussed in detail in M17.

The use of a photometric prior typically reduces the difference between photometric and spectroscopic redshifts, whilst also reducing the fraction of catastrophic failures. We find that the COSMOS region provides the most accurate photometric redshifts compared to spectroscopic redshifts.  The GAMA region has a 97\% completeness fraction, which creates an unbiased sample and is arguably a better indicator of photometric redshift efficacy (see M17).  

    \begin{table}
\begin{minipage}{0.475\textwidth}
\centering
\caption{Our photometric redshift quality, with and without priors, as discussed in \S \ref{subsec:photozs}. In each our  fields we list below: the number of secure spectroscopic redshifts used in the comparisons ($N_s$), the normalised median absolute deviation ($\sigma_{_\text{NMAD}}$), mean $|\Delta{z}| / (1+z_s)$, average bias $\Delta{z} = (z_\text{spec} - z_\text{phot})$, and fraction of catastrophic outliers ($\eta_{_1}$ and $\eta_{_2}$) defined in two ways.}
\label{tab:photo-z-results}
\begin{tabular}{@{}lcccccc} 
\hline
Field	&	$N_\text{s}$ & $\sigma_{_\text{NMAD}}$	&	$\frac{|\Delta{z}|}{(1+z_\text{s})}$	&	$\Delta{z}$ & $\eta_{_1}$\footnote{Catastrophic outliers determined as $|\Delta{z}|/(1+z_\text{spec}) > 0.15$.} & $\eta_{_2}$\footnote{Catastrophic outliers determined as $|\Delta{z}|/(1+z_\text{spec}) > 3 \times \sigma_{_\text{NMAD}}$.} \\
\hline
\multicolumn{7}{c}{With a magnitude prior} \\
\hline
UDS & 2648 & 0.053 & 0.045 & 0.01 & 5.3\% & 5.0\% \\
VIDEO & 4382 & 0.044 & 0.038 & 0.01 & 2.9\% & 3.3\% \\
COSMOS & 5467 & 0.013 & 0.010 & 0.00 & 0.5\% & 2.5\% \\
GAMA & 55199 & 0.049 & 0.044 & -0.02 & 2.4\% & 2.5\% \\
\hline
\multicolumn{7}{c}{Without a magnitude prior} \\
\hline
UDS & 2648 & 0.051 & 0.045 & 0.01 & 5.3\% & 5.3\% \\
VIDEO & 4382 & 0.048 & 0.042 & 0.02& 3.4\% & 3.5\% \\
COSMOS & 5467 & 0.013 & 0.011 & 0.00 & 0.5\% & 3.2\% \\
GAMA & 55199 & 0.060 & 0.052 & -0.03 & 3.4\% & 1.7\% \\
\hline
\end{tabular}
\end{minipage}
\end{table}

\begin{figure}
    \includegraphics[width=3.5in]{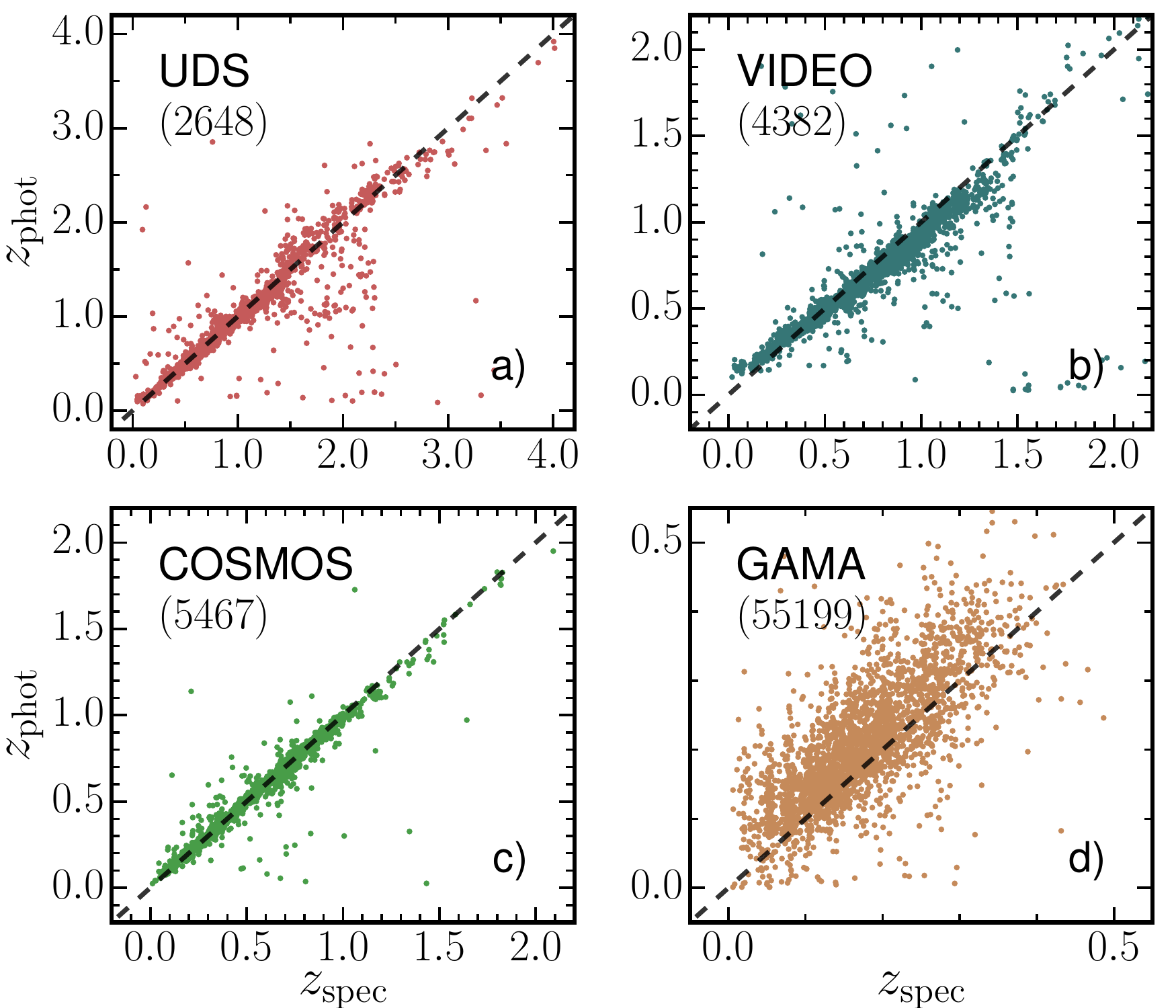}
	\caption{The comparison between best-fit photometrically derived redshifts, $z_\mathrm{phot}$, and spectroscopically measured redshifts, $z_\mathrm{spec}$, in the a) UDS, b) VIDEO, c) COSMOS, and d) GAMA regions, similar to that shown in M17. Shown within parenthesis are the number of science-quality spectroscopic redshifts within each field which we use for calibrating the photometric redshifts.  The normalised median absolute deviation, average offset and outlier fraction of our photometric redshifts are discussed in more detail in M17. }
    \label{fig:photoz}
\end{figure}

The corrections described above results in PDFs which accurately represent the probability of every galaxy at every redshift over the range $0 < z < 6$. The integral of the PDF over some redshift range measures the probability of the galaxy being found within each redshift range.

\subsection{Stellar mass estimates}

We calculate stellar masses using {\tt{smpy}}, a custom spectral energy distribution (SED) fitting code, first introduced in \citet{Duncan2014} and available online\footnote{https://www.github.com/dunkenj/smpy/}.  To calculate stellar masses we use the \citet[][BC03]{Bruzual2003} stellar population synthesis models with a \citet{Chabrier2003} initial mass function (IMF).   The model ages we use are varied between 0.01--13.7 Gyr, as in M17.   We use simple star formation history $\tau$-models which are  exponentially increasing or decreasing, or constant, with values of $|\tau|$ between 0.01--13.7 Gyr. We also include an option for a constant star-formation history within these models. We use a \citet{Calzetti2000} model for dust extinction, with ($A_V$) allowed to vary between 0 -- 4 magnitudes. The stellar metallicity varies between the range $0.005 < Z/Z_\odot < 2.5$.   

As in M17 the stellar mass is estimated at all redshifts in the photometric redshift fitting range simultaneously. The resulting likelihood-weighted mean is then defined as:

\begin{equation}
\mathcal{M}_*(z) = \frac{\sum_t w_t(z)\ \mathcal{M}_{*,t}(z)}{\sum_t w_t(z)}
\end{equation}

\noindent where we sum over all galaxy template types, $t$, with ages less than the age of the Universe at the redshift $z$ and $\mathcal{M}_{*,t}$ is the best-fit stellar mass for each galaxy template. The likelihood, $w_t(z)$, is determined by

\begin{equation}
w_t(z) = \exp\left( -\chi^2_t(z) / 2 \right)
\end{equation}
where $\chi^2_t(z)$ is given by
\begin{equation}
\chi^2_t(z) = \sum_j \frac{(F_{j,t}(z) - F_j^\text{obs})^2}{\sigma_j^2}.
\end{equation}
We take the sum over the $j$ available filters for each particular galaxy. This includes the observed photometric fluxes, $F_j^\text{obs}$, as well as  the photometric flux error, $\sigma_j$.  We fit our photometry to a library of 34,803 synthetic SEDs to achieve this. Stellar mass as a function of redshift within each region is plotted and discussed in M17. 

%stopped may 19 '22

\subsection{Merger Fraction Measurements}

The method for measuring merger fractions in this paper is essentially the same as in M17.  We give a summary of that process here, although see M17 for a more detailed explanation for how galaxy pairs are determined based on the data. 

\subsubsection{Formalism and Basic Selection of Pairs}

The aim of any close-pairs statistics study is to measure the fraction of galaxies undergoing a merger within a defined sample. Spectroscopic studies performed in the local Universe often define a close-pair as two galaxies within some \textit{projected} separation and within some relative velocity offset, typically taken to be $\Delta v \leq 500$ km s$^{-1}$ \citep[e.g.,][]{Bluck2012, Tasca2014}. This definition, or one similar to it, can then be used to measure the pair fraction, $f_\text{pair}$, defined as:

\begin{equation}
	f_\text{pair} (z, {\rm selection}, \mu) = (N_\mathrm{pairs} / N_\mathrm{tot}),
\end{equation}

\noindent where $N_\mathrm{pairs}$ is the number of galaxy close-pairs and $N_\mathrm{tot}$ is the total number of galaxies in the parent sample. The former number is the number of close-pairs rather than the number of galaxies in close-pairs. The galaxy merger fraction, or the fraction of galaxies in pairs, would be a factor of two larger \cite[][]{Conselice2014a}. Note that the parents sample and the galaxies from which the pairs are taken are defined by a redshift (z) and selection, either stellar mass or number density limit, as well as a pair ratio ($\mu$) range. 

We use the method described in M17 to find pairs, which enables measurements of the merger fraction for a flux-limited photometric samples of galaxies. To do this we use our photometric redshift ($z_\text{phot}$) probability distribution functions (PDFs), which accounts for the uncertainty in redshift during the pair selection procedure.

To start our selection of close pairs, an initial list of candidate projected close-pairs is constructed from the science catalogues in each region described in Section~2. This is achieved by selecting pairs with a projected separation less than the maximum angular separation of the redshift range probed and corresponding to a 30 kpc upper limit separation.

\subsubsection{The pair probability function}

A redshift probability function, $\mathcal{Z}(z)$, is calculated for each candidate close-pair system.   This includes the number of close-pairs which contribute to this based on the line-of-sight information encoded within each galaxy's redshift PDFs. This quantity is defined as:

\begin{equation}
	\label{eqn:pyrus:zpf}
	\mathcal{Z}(z) = \frac{2 \times P_1(z) \times P_2(z)}{P_1(z) + P_2(z)} = \frac{P_1(z) \times P_2(z)}{N(z)}
\end{equation}

\noindent where $P_1(z)$ and $P_2(z)$ are the PDFs for the primary (defined as the more massive galaxy) and secondary galaxies, respectively. Within this equation, the redshift probability function is normalised such that each pairing can maximally contribute a single pair when integrated over the full redshift range. This can be expressed as:
\begin{equation}
	N_{\mathrm{pair},j} = \int_{0}^{\infty} \mathcal{Z}_j(z) \mathrm{d}z
\end{equation}
which ranges between 0 and 1. As each galaxy in the primary sample is allowed to have multiple companions, each \textit{projected} close-pair is considered separately and included in the total integrated pair count.

\subsubsection{Close-pair constraints}

We use redshift dependent masks for the remaining close-pair selection criteria. This is zero where conditions are not met and unity otherwise.  The angular separation mask, $\mathcal{M}^\theta(z)$, is defined as:
\begin{equation}
	\mathcal{M}^\theta(z) = 
    \begin{cases}
    1, & \text{if}\ \theta_\mathrm{min}(z) \leq \theta_j \leq \theta_\mathrm{max}(z), \\
    0, & \text{otherwise.}
    \end{cases}
\end{equation}
Where the minimum and maximum angular separations are defined as $\theta_\text{min} = r_p^\text{min}/d_A(z)$ and $\theta_\text{max} = r_p^\text{max}/d_A(z)$, respectively, as a function of redshift, and where $d_A(z)$ is the angular diameter distance.     

Close galaxy pairs are also selected based on stellar mass and therefore the pair selection mask, $\mathcal{M}^\text{pair}$, is defined as:

\begin{equation}
\label{eqn:pyrus:pair-mask}
	\mathcal{M}^\text{pair} = 
    \begin{cases}
	1, & \text{if}\ \mathrm{M}_*^{\text{lim},1}(z) \leq \mathrm{M}_{*,1}(z) \leq \mathrm{M}_{*,\text{max}} \\
       & \text{and}\ \mathrm{M}_*^{\text{lim},2}(z) \leq \mathrm{M}_{*,2}(z) \\
    0, & \text{otherwise.}
	\end{cases}
\end{equation}

\noindent where $\mathrm{M}_{*,1}(z)$ and $\mathrm{M}_{*,2}(z)$ are the stellar masses as a function of redshift for the primary and secondary galaxies, respectively. The limiting stellar masses in Equation \ref{eqn:pyrus:pair-mask} are given by
\begin{equation}
	\mathrm{M}_*^{\text{lim},1}(z) = \text{max}(\mathrm{M}_*^\text{min}, \mathrm{M}_*^\text{comp}(z))
\end{equation}
and
\begin{equation}
	\mathrm{M}_*^{\text{lim},2}(z) = \text{max}(\mu \mathrm{M}_{*,1}(z), \mathrm{M}_*^\text{comp}(z)).
\end{equation}
The value $\mathrm{M}_*^\text{comp}(z)$ represents the redshift-dependent stellar mass completeness limit of each survey region, whereas $\mathrm{M}_*^\text{min}$ is the minimum stellar mass of the primary sample selection. The value of $\mu$ is the stellar mass ratio, where we use the limits $\mu = 1/4$ for major mergers, and $\mu = 1/10$ for minor mergers. The redshift-dependent stellar mass completeness limits are discussed in M17. Application of the pair selection mask in Equation \ref{eqn:pyrus:pair-mask} ensures that (1) the primary galaxy is within the correct stellar mass range, (2) the stellar mass ratio of the primary and secondary galaxies corresponds to either major or minor mergers, and (3) primary and secondary galaxies are both above the stellar mass completeness limit of the survey region they are found. With these masks for each \textit{projected} close-pair, the pair probability function, $\text{PPF}(z)$, is simply given by
\begin{equation}
\label{eqn:pyrus:ppf}
	\text{PPF}(z) = \mathcal{Z}(z) \times \mathcal{M}^\theta(z) \times \mathcal{M}^\text{pair}(z).
\end{equation}
The integral of the PPF provides the \textit{unweighted} number of close-pairs (as defined by the chosen selection criteria) that two galaxies contribute to the measured pair fraction.

\subsection{Selection effect corrections}
\label{mergers:counting-pairs:corrections}

The flux-limited nature of the photometric surveys used in this study combined with the stellar mass selection of close-pairs requires several selection effects to be appropriately accounted for.

Firstly, a primary galaxy may possess a stellar mass close to the stellar mass completeness limit of the survey region at some redshift. Such a scenario may reduce the stellar mass range in which secondary companions can be found and thus result in fewer companions found. To address this bias, a statistical correction is made using the galaxy stellar mass function, $\phi(\mathrm{M}_*, z)$, at the redshift range of interest. Simply, each secondary companion of a primary galaxy is assigned a weighting defined as 

\begin{equation}
	\omega_2^\text{comp}(z) = {\Bigg[ \frac{\int_{\mathrm{M}_*^{\text{lim}}(z)}^{\mathrm{M}_{*,1}(z)} \phi(\mathrm{M}_*, z)\ \mathrm{d}\mathrm{M}_*}{\int_{\mu \mathrm{M}_{*,1}(z)}^{\mathrm{M}_*^1(z)} \phi(\mathrm{M}_*, z)\ \mathrm{d}\mathrm{M}_*} \Bigg]}^{-1}.
\end{equation}

\noindent This correction is essentially the inverse of the ratio of galaxy number densities above and below the stellar mass completeness limit. Applying this correction provides close-pair fractions corresponding to a volume-limited study. These weights are an analogue of the luminosity weights presented in \citet{Patton2000}.

A second weighting is applied to the primary galaxies in order to correct for those objects that will have fewer \textit{observed} companions because of their proximity to the completeness limit of the survey region. The primary galaxy completeness weight is defined as 

\begin{equation}
	\omega_1^\text{comp}(z) = \frac{\int_{\mathrm{M}_*^{\text{lim}}(z)}^{\mathrm{M}_*^\text{max}(z)} \phi(\mathrm{M}_*, z)\ \mathrm{d}\mathrm{M}_*}{\int_{\mathrm{M}_*^\text{min}(z)}^{\mathrm{M}_*^\text{max}(z)} \phi(\mathrm{M}_*, z)\ \mathrm{d}\mathrm{M}_*}
\end{equation}

\noindent where $\mathrm{M}_*^\text{min}$ and $\mathrm{M}_*^\text{max}$ are the minimum and maximum stellar masses of the primary sample for which the merger fraction is being calculated. 

Galaxies close to the survey edges or near `bad' areas with corrupt photometry (e.g. bright stars, cross-talk) may possess a reduced spatial area in which to find companions.   The spatial search area is a function of a fixed physical search radius, thus the correction is a function of redshift. We produce a mask image which is 1 where good photometry exists and 0 elsewhere. For each galaxy within the primary sample `photometry' is performed on this mask image over the spatial search area. An area weight is assigned, defined as
\begin{equation}
	\omega_\text{area}(z) = \frac{1}{f_\text{area}(z)},
\end{equation}
where $f_\text{area}(z)$ is the fraction of the mask image with good photometry within the search annulus.  Since the search area is a function of redshift, thus so too are the area weights.

The final weighting applied is based on the photometric redshift quality, encoded by the Odds sampling rate. The Odds parameter, $\mathcal{O}$, is defined by \citet{Benitez2000} and \citet{Molino2014} as 
\begin{equation}
	\mathcal{O}_j = \int_{-K(1+z_p)}^{+K(1+z_p)} P_j(z)
\end{equation}
for each galaxy $j$.  In this equation $z_p$ is the galaxy's best-fit photometric redshift, and $P_j(z)$ is the redshift PDF of the galaxy. The value of $K$ is the typical photometric redshift accuracy of the data. For example, \citet{Molino2014} use $K = 0.0125$  from 20 medium-band ($\sim 300$\AA\ in width) filters. In this work however, we typically use broad-band filters, and thus a larger value of $K = 0.05$ is chosen as a result of comparing photometric and spectroscopic redshifts.

The Odds sampling rate (OSR) is defined as the fraction of galaxies with an Odds parameter above this criteria, normalised by the total number of galaxies as a function of apparent magnitude, $m$. 
\begin{equation}
\text{OSR}(m) = \frac{\sum N(\mathcal{O} \geq 0.3)}{\sum N(\mathcal{O} \geq 0)},
\end{equation}
which is used to compute the weight for a particular galaxy, $j$, with magnitude $m_j$ as
\begin{equation}
\omega_j^\text{OSR} = \frac{1}{\text{OSR}(m_j)}.
\end{equation}

\noindent Combining these weights, the primary and secondary galaxy weights can be consolidated and defined. Each secondary galaxy around a primary galaxy is assigned a weight given by

\begin{equation}
	\omega_2(z) = \omega_1^\text{area}(z) \times \omega_1^\text{comp}(z) \times \omega_2^\text{comp}(z) \times \omega_1^\text{OSR} \times \omega_2^\text{OSR}
\end{equation}
and each primary galaxy is weighted by 
\begin{equation}
	\omega_1(z) = \omega_1^\text{comp}(z) \times \omega_1^\text{OSR}.
\end{equation}

We use these weights to determine the pair and merger fractions which are used throughout the analyses of this paper.

\section{Results}
\label{sec:results}
\subsection{Merger Fractions}

The close-pair fraction is defined as the number of galaxy close-pairs divided by the total number of galaxies in the primary sample. This is, for discrete measurements, given as: $f_\text{pair} = N_\text{pair} / N_\text{tot}$. In our analysis the number of close-pairs associated with galaxy $i$ in the primary sample, over the redshift range $z_\text{min} < z < z_\text{max}$, is given by
\begin{equation}
\label{eqn:pyrus:N2}
	N_\text{pair}^i = \sum_j \int_{z_\text{min}}^{z_\text{max}} \omega_2^j(z) \times \mathrm{PPF}_{i,j}(z)\ \mathrm{d}z
\end{equation}
where $j$ indexes all potential secondary galaxies around the primary galaxy and $\mathrm{PPF}_{i,j}(z)$ are the pairs' pair probability function. Accordingly, the contribution from the primary galaxy over the same redshift range is given by
\begin{equation}
\label{eqn:pyrus:N1}
	N_1^i = \sum_i \int_{z_\text{min}}^{z_\text{max}} \omega_1^i(z) \times P_i(z) \times \mathcal{S}_1^i(z)\ \mathrm{d}z
\end{equation}
where $\omega_1^i(z)$ is its weighting, $\mathcal{S}_1^i(z)$ is the primary galaxy selection function and $P_i(z)$ its normalised redshift probability density function as described in \S \ref{sec:data_products}.  When the primary galaxy has a stellar mass and redshift PDF contained within our limits then $N_1^i = \omega_1^i$, and is always equal to or greater than unity.

The statistical weighting of the close-pair fraction is measured by summing Equations \ref{eqn:pyrus:N2} and \ref{eqn:pyrus:N1} over all primary galaxies. In the redshift range $z_\text{min} < z < z_\text{max}$ the close-pair fraction is then given by
\begin{equation}
\label{eqn:pyrus:fm}
	f_\text{pair} = \frac{\sum_i N_\text{pair}^i}{\sum_i N_1^i}.
\end{equation}

We thus utilize this approach for our measurements of the pair fractions for the total, major, and minor mergers throughout this paper.

\subsection{Merger Fraction Evolution}
\label{subsec:mergerfraction_evo}

The first observation we describe is what we call the ``total'' merger fraction evolution.  This has to be carefully done, as we are measuring down to our mass limit the {\em total} number of mergers which have a mass ratio $\mu > 1/10$, which includes both major and minor mergers.  We later deconstruct this using the observed major mergers with mass ratios of 1/4 to obtain the minor merger fraction, which are those mergers with mass ratios of 1/4 down to 1/10.  The following sections describe this data and processes.

\subsubsection{Total merger pair fractions}
\label{minor:fractions}

First, we describe our basic observation of the total fraction of mergers down to mass ratios of 1:10. This is the limit for how deep we can observe minor mergers for our massive galaxies.  We measure the total (major plus minor) merger pair fractions  using the same method as outlined in \S \ref{sec:data_products}.  We later examine the minor mergers from this total merger fraction evolution. 

Pair fractions are presented for the same overall galaxy selections as previously described in M17, however only fractions at 5--30kpc are reported for the sake of space and improved number statistics over a 20 kpc upper limit. Pair fractions for all mergers with mass ratios greater than this within each survey region are listed in Table~1 for M$_{*} > 10^{11}$ \solm galaxies, and Table~2 for galaxies selected at a constant number density.

\begin{figure*}
    \vspace{-1cm}
    \begin{center}
	\includegraphics[width=5in]{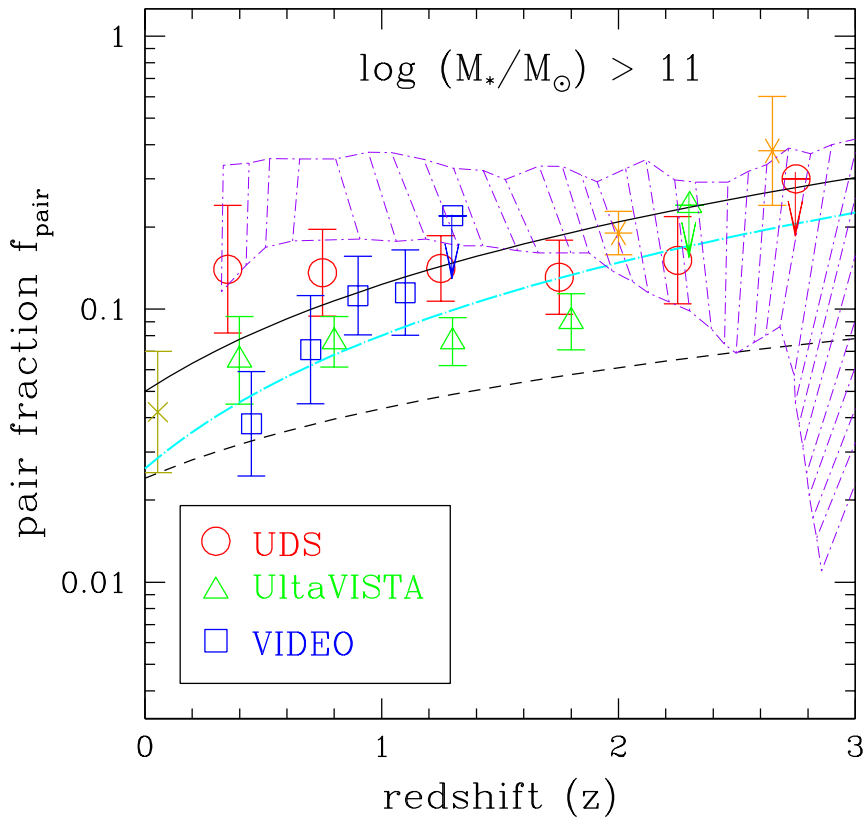}
	\end{center}
    \label{fig:fm-11}
    \vspace{-3.5cm}
    \caption{The total (see text) measured pair fraction, $f_\text{pair}$, for galaxies at $M_* > 10^{11} M_\odot$ down to a stellar mass ratio of 1:10. Measurements are made in the UDS (red circles), UltraVISTA (green triangles), VIDEO (blue squares) and GAMA (gold cross) survey fields. A power-law (black solid line) is fit to the data points with best fitting parameters described in the text.   The lower dashed line shows the major merger pair fraction for the same stellar mass range from M17.   The two orange points at higher redshifts are the merger fractions for minor mergers from Bluck et al. (2012).  The purple shaded region shows the 1$\sigma$ variation in the pair fraction as measured by using 24 light cones based on the H15 semi-analytical model.  The solid line is the best fit line for the total merger history, the dashed line for the major mergers, and the minor mergers are shown by the cyan dot-dash line.}
    \vspace{1cm}
\end{figure*}

We measure total pair fractions for massive galaxies at $\mathcal{M}_* >10^{11}\ \mathrm{M}_\odot$ out to $z \sim 2.25$, beyond which measurements of the pair fraction are upper limits due to the stellar mass completeness limits of our fields. Studying mergers at higher redshifts will require substantially deeper data over large fields, which will be possible with upcoming surveys such as Euclid and Rubin/LSST. The values that we calculate in this paper are thus the best that can be done for some time.

The total merger fraction evolution results for our sample are plotted in Figure~2 for the GAMA (gold cross), UKIDSS UDS (open red circles), VIDEO (open blue squares), and COSMOS (open green triangles) regions.  In this plot we also show the major merger fraction as a dashed line to compare with.

Generally, we find that the value of total $f_\text{pair}$ is observed to increase steadily from $\sim3.5\%$ at $z=0$ to $\sim15\%$ at $z=2.25$, however differences are seen between fields. Fractions within the UKIDSS UDS region remain approximately constant at $0.35 < z < 2.25$, with $f_\text{total} \approx 0.13$, while measurements in the VIDEO region rise sharply from 4\% to 12\% at $z\sim0.35$ and $z\sim1$, respectively. A similar evolution is seen in the Ultra-VISTA region where $f_\text{total} \approx 0.07$ at $z < 1.5$. These differences are largely due to cosmic variance and to the fact that different inherent environments of the galaxy population are being probed within the different fields.  

The pair fractions for the local universe are calculated using the GAMA low redshift data utilizing both photometric and spectroscopic redshifts (assuming a $\Delta{v} < 500$ km/s difference for the spectroscopic sample). These are shown in all plots as gold crosses at the lowest redshifts. Similarly as when probing major mergers, we find an excellent agreement in the pair fractions measured in these two different ways. A GAMA spectroscopic pair fraction at $0.005 < z < 0.2$ of $0.042\pm0.028$ is in agreement with the fractions measured photometrically at $\sim4\%$. 

Following the same procedure adopted in M17, pair fractions are also measured within 24 light cones extracted from the \citet{Henriques2014} semi-analytic model. The model is treated as a complete spectroscopic sample, maintaining the same velocity difference criteria as used within the GAMA region. The 1$\sigma$ uncertainty region from the model is shown as the light purple shaded area in Figure~2. This model predicts pair fractions that remain approximately constant at $\sim20\%$ with redshift. This qualitative evolution is consistent with the UKIDSS UDS and COSMOS regions, albeit systematically larger by a factor of 2--3 relative to the observations. 

As in M17, the pair fractions are fitted with a power-law of the form $f_\text{pair}(z) = f_0(1+z)^m$.  Uncertainties on these parameters are estimated by sampling individual data points from a normal distribution, centred on $f_\text{pair}$ with the standard deviation given by the errors quoted in Table~1, and building a distribution of parameters by performing a least-squares fitting routine $10^4$ times. The best-fit parameters (Table~4) are:

%june 30, 2021 - edited equation.
\begin{equation}
f_{\rm total}(z, M_{*} > 11) = 0.051^{+0.009}_{-0.008} \times (1+z)^{1.30^{+0.32}_{-0.33}}.
\end{equation}

\noindent This fit is plotted as a solid line in Figure~2.  Also shown in  Figure~2 is the best-fit major merger parametrisation for the same massive sample of galaxies (Table~2; M17), given by the dashed curve. It is apparent that the evolution of major and total mergers shares a similar slope over this redshift range, and that the total merger pair fraction is consistently a factor of $\sim2-3$ larger than the major merger pair fraction.

\begin{table*}
\centering
\caption{Measured total pair fractions and error analysis for a mass-selected sample of $M/M_\odot > 11$ within the UDS, UltraVISTA, VIDEO and GAMA survey fields down to a merger ratio of 1:10. The redshift bin edges ($z_\text{low}$ and $z_\text{high}$), number of merging pairs ($N_\text{pair}$), total number of galaxies in the primary sample ($N_\text{pri}$), Poisson error contribution ($\Delta_\text{poisson}$), cosmic variance error contribution ($\Delta_\text{CV}$), measured pair fraction ($f_\text{pair}$) and the total error on the pair fraction ($\Delta f_\text{pair}$). Measurements in parentheses are not mass complete and are plotted as 3$\sigma$ upper limits.}
\label{table:fm-11}
\begin{tabular}{lrrrrrrrrrrll}
\hline
\hline
$z_\text{low}$ & $z_\text{high}$ & $N_\text{pair}$ & $\Delta N_\text{pair}$ & $N_\text{pri}$ & $\Delta N_\text{pri}$ & $\Delta_\text{boot}$ & $\Delta_\text{poisson}$ & $\Delta_\text{CV}$ &  $f_\text{pair}$ & $\Delta f_\text{pair}$ \\
\hline
\multicolumn{11}{c}{UDS} \\
\hline
0.2 &     0.5 &  8.013 &  2.541 &  57.305 &  6.870 & 0.041 &      0.077 & 0.049 &  0.140 & 0.100 \\
0.5 &     1.0 & 40.538 & 11.348 & 297.605 & 15.416 & 0.038 &      0.033 & 0.033 &  0.136 & 0.060 \\
1.0 &     1.5 & 69.701 &  8.037 & 493.847 & 20.119 & 0.015 &      0.026 & 0.034 &  0.141 & 0.045 \\
1.5 &     2.0 & 54.421 &  7.469 & 415.393 & 16.869 & 0.016 &      0.027 & 0.037 &  0.131 & 0.048 \\
2.0 &     2.5 & 33.632 &  4.565 & 223.385 & 12.493 & 0.018 &      0.040 & 0.051 &  0.151 & 0.067 \\
2.5 &     3.0 & 22.775 &  3.593 & 205.165 &  9.842 & 0.017 &      0.035 & 0.048 &  (0.30) & - \\
\hline
\multicolumn{11}{c}{UltraVISTA} \\
\hline
0.2 &     0.5 &  37.713 &  6.060 &  581.772 & 24.366 & 0.010 &      0.016 & 0.023 &  0.065 & 0.029 \\
0.5 &     1.0 & 167.173 & 12.440 & 2198.231 & 45.986 & 0.005 &      0.009 & 0.014 &  0.076 & 0.018 \\
1.0 &     1.5 & 128.718 &  8.976 & 1686.829 & 35.784 & 0.005 &      0.010 & 0.013 &  0.076 & 0.017 \\
1.5 &     2.0 &  88.148 &  6.827 &  978.633 & 26.957 & 0.006 &      0.014 & 0.018 &  0.090 & 0.024 \\
2.0 &     2.5 &  73.502 &  7.166 &  733.980 & 21.523 & 0.010 &      0.017 & 0.024 &  (0.24) & - \\
\hline
\multicolumn{11}{c}{VIDEO} \\
\hline
0.2 &     0.5 & 10.877 &  2.028 & 285.713 & 14.582 & 0.007 &      0.017 & 0.011 &  0.038 & 0.021 \\
0.5 &     0.7 & 11.325 &  2.385 & 159.533 & 10.072 & 0.014 &      0.031 & 0.021 &  0.071 & 0.041 \\
0.7 &     0.9 & 34.655 &  4.272 & 307.892 & 14.873 & 0.012 &      0.029 & 0.031 &  0.112 & 0.044 \\
0.9 &     1.1 & 24.191 &  3.474 & 210.865 &  9.861 & 0.015 &      0.035 & 0.032 &  0.115 & 0.050 \\
1.1 &     1.3 & 18.897 &  2.672 & 235.957 & 12.154 & 0.011 &      0.027 & 0.022 &  (0.22) & - \\
\hline
\multicolumn{11}{c}{GAMA} \\
\hline
0.01 &     0.1 &  6.208 & 1.470 &  147.706 &  9.117 & 0.010 &      0.025 & 0.009 &  0.042 & 0.028 \\
\hline
\hline
\end{tabular}
\end{table*}

\begin{table*}
\centering
\caption{Measured pair fractions and error analysis for a number density selected sample of $n = 1\times10^{-4} \textrm{Mpc}^{-3}$ within the UDS, UltraVISTA, VIDEO and GAMA survey fields down to a merger ratio of 1:10. The redshift bin edges ($z_\text{low}$ and $z_\text{high}$), number of merging pairs ($N_\text{pair}$), total number of galaxies in the primary sample ($N_\text{pri}$), Poisson error contribution ($\Delta_\text{poisson}$), cosmic variance error contribution ($\Delta_\text{CV}$), measured total pair fraction ($f_\text{pair}$) and the total error on the pair fraction ($\Delta f_\text{pair}$). Measurements in parentheses are not mass complete and are plotted as 3$\sigma$ upper limits.}
\label{table:fm-1em4}
\begin{tabular}{llrrrrrrrrrll}
\hline
\hline
$z_\text{low}$ & $z_\text{high}$ & $N_\text{pair}$ & $\Delta N_\text{pair}$ & $N_\text{pri}$ & $\Delta N_\text{pri}$ & $\Delta_\text{boot}$ & $\Delta_\text{poisson}$ & $\Delta_\text{CV}$ &  $f_\text{pair}$ & $\Delta_\text{tot}$ \\
\hline
\multicolumn{11}{c}{UDS} \\
\hline
0.2 &     0.5 &  0.129 & 0.088 &   7.526 &  2.469 & 0.014 &      0.073 & 0.006 & 0.018 &  0.075 \\
0.5 &     1.0 &  6.126 & 1.761 &  80.644 &  6.783 & 0.021 &      0.045 & 0.018 & 0.075 &  0.053 \\
1.0 &     1.5 & 48.454 & 6.557 & 355.353 & 17.034 & 0.017 &      0.030 & 0.033 & 0.137 &  0.048 \\
1.5 &     2.0 & 64.319 & 7.259 & 480.944 & 16.441 & 0.015 &      0.025 & 0.038 & 0.135 &  0.048 \\
2.0 &     2.5 & 61.488 & 5.683 & 436.614 & 17.639 & 0.012 &      0.027 & 0.048 & (0.142) &  (0.057) \\
2.5 &     3.0 & 88.770 & 9.065 & 551.150 & 16.377 & 0.015 &      0.026 & 0.070 & (0.163) &  (0.077) \\
\hline
\multicolumn{11}{c}{UltraVISTA} \\
\hline
0.2 &     0.5 &   9.338 &  3.020 &   88.899 &  8.897 & 0.032 &      0.051 & 0.037 & 0.105 &  0.070 \\
0.5 &     1.0 &  68.836 &  8.151 &  787.616 & 26.284 & 0.010 &      0.015 & 0.016 & 0.086 &  0.024 \\
1.0 &     1.5 &  97.903 &  7.854 & 1265.407 & 33.127 & 0.006 &      0.012 & 0.013 & 0.078 &  0.018 \\
1.5 &     2.0 &  98.190 &  6.742 & 1129.751 & 27.895 & 0.006 &      0.013 & 0.017 & (0.086) &  (0.022) \\
\hline
\multicolumn{11}{c}{VIDEO} \\
\hline
0.2 &     0.5 &  1.583 &  0.710 &  80.000 &  8.332 & 0.009 &      0.022 & 0.006 & 0.019 &  0.024 \\
0.5 &     0.7 &  4.241 &  1.531 &  56.047 &  5.666 & 0.027 &      0.053 & 0.022 & 0.075 &  0.064 \\
0.7 &     0.9 &  9.335 &  2.163 &  84.671 &  7.293 & 0.025 &      0.054 & 0.031 & 0.110 &  0.067 \\
0.9 &     1.1 & 13.965 &  3.109 & 107.491 &  7.789 & 0.026 &      0.053 & 0.037 & 0.132 &  0.070 \\ 
1.1 &     1.3 & 12.657 &  2.482 & 167.980 & 10.941 & 0.014 &      0.032 & 0.022 & (0.077) &  (0.041) \\
1.3 &     1.5 & 25.295 &  6.938 & 194.760 & 10.732 & 0.035 &      0.038 & 0.036 & (0.128) &  (0.063) \\
\hline
\multicolumn{11}{c}{GAMA} \\
\hline
0.01 &     0.1 &  1.722 & 0.741 &  60.046 &  6.271& 0.011 &      0.032 & 0.006 & 0.029 &  0.034 \\
\hline
\hline
\end{tabular}
\end{table*}

\subsubsection{The minor merger pair fraction}

With measurements of the major merger and `total' merger pair fraction in hand, the signal from minor mergers can be extracted.   We do this first by inferring the minor merger pair fraction evolution, or $f_\text{minor} (z)$.   From this we later derive the minor merger rate. Ultimately, the fraction of minor merger pairs is the residual of subtracting the number of major mergers from the total number of mergers we measure using our methodology.  In numerical terms this is simply:

\begin{equation}
f_\text{minor} = (f_\text{pair} - f_\text{major})
\end{equation}

\noindent This subtraction can be achieved in two ways: using the raw counts of primary galaxies and close pairs in each redshift bin, or taking the pair fraction fits and subtracting the major from total pair fraction fits. The latter is the most direct method of achieving this, and can account for field to field variations in this history.  We thus use this method henceforth.  However, the direct counting method gives the same results.

%For completeness, total, major, and minor pair fractions are presented in their raw form in Appendix \ref{apdx:minormerger}. 

\begin{figure}
    \vspace{-1cm}
    \begin{center}
	\includegraphics[width=3.5in]{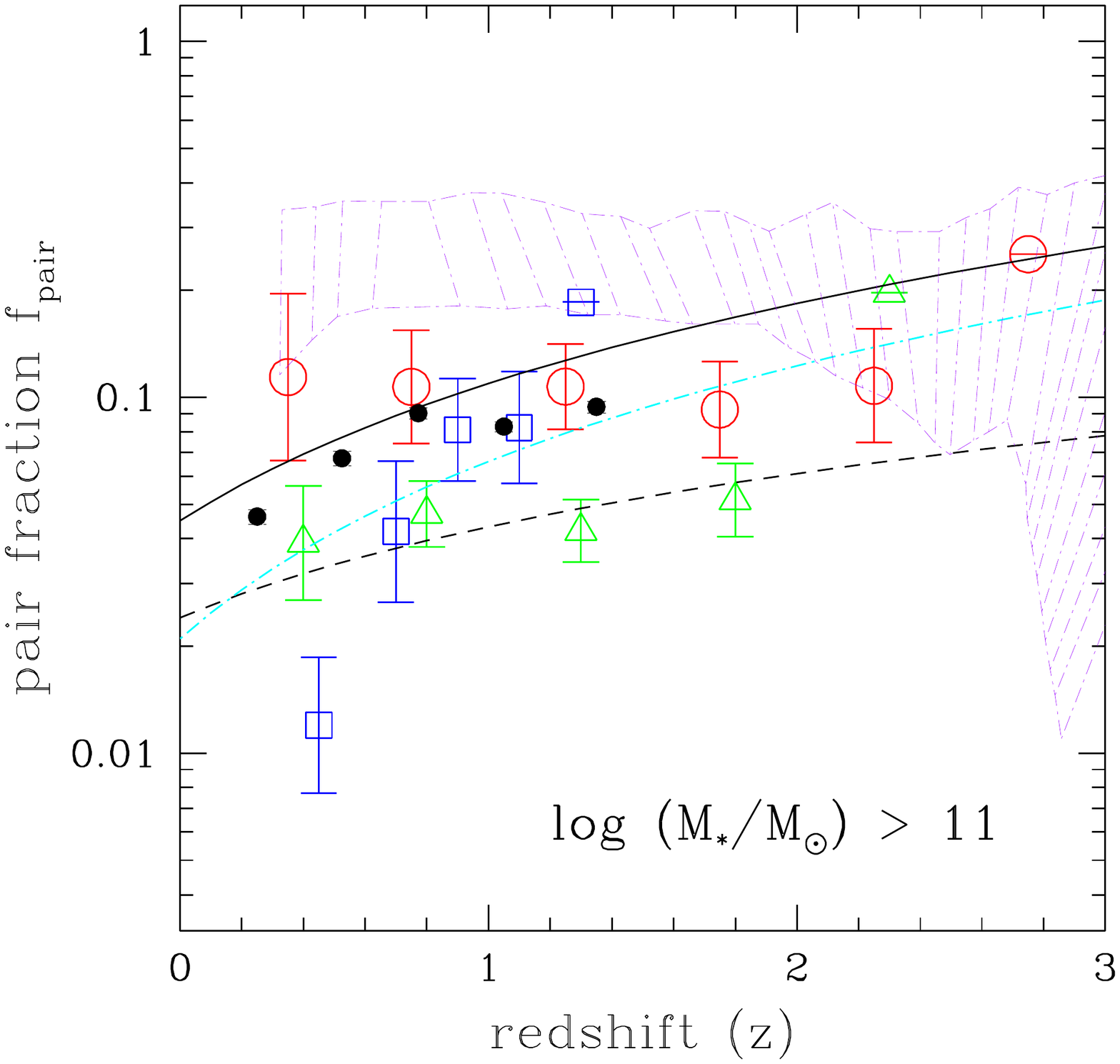}
	\end{center}
    \label{fig:fm-11}
    \vspace{-3cm}
    \caption{Comparison between the total merger pair fraction history (solid line) along with the major mergers (dashed line), and derived minor mergers (dot-dashed cyan line).  These are from the best fits to these quantities, showing that minor mergers are more common up to $z \sim 3$ than major mergers are. We furthermore plot the \citep[][]{Man2016A} results for measuring the minor merger fraction from CANDELS.  Also shown are the model predictions as discussed in Figure~2. }
    \vspace{1cm}
\end{figure}

We make many realizations of the total and major pair fractions using the best-fit parametrisations and associated uncertainties given in Tables~1\&2. This includes taking asymmetric error distributions into account. Within each realisation, the major merger pair fraction is subtracted from the total pair fraction at $0 < z < 3.5$.  The residual of this is then fit with the same power-law form as used previously.  After collecting the distributions of fitting parameters we calculate the realisations for the 50th, 16th and 84th percentiles on each parameter. These best-fit parameters and the 1$\sigma$ uncertainties of the minor merger pair fraction are discussed below and in Table~4.

Figure~3 displays the evolution of the total (major + minor), major and minor pair fractions \citep[][]{Man2016A}.  Averaged over $z<2$, the major and minor pair fractions are approximately equal at low-z.  However, there are more minor mergers at $z > 0.5$.  If the minor merger pair fraction evolves as predicted beyond $z=2$, it would become a factor of two larger than the major merger pair fraction by $z=3$.  We fit the minor merger fraction as a function of redshift as:

\begin{equation}
f_{\rm minor}(z, M_{*} > 11) = 0.023^{+0.008}_{-0.007} \times (1+z)^{1.52^{+0.30}_{-0.33}}.
\end{equation}

\noindent Note that we find that the intermediate mass ($\mathcal{M}_* > 10^{10}\ \mathrm{M}_\odot$) galaxies have a strong evolution in the minor merger pair fraction, however it is likely this is due to the relatively strict redshift range ($z < 1$) in which this quantity is measured.   As with the larger mass selection, the major and minor merger pair fractions are approximately comparable at this redshift regime.

\subsection{Merger Time-Scales}
\label{subsec:merger_timescales}
Ultimately, we want to determine the role of minor mergers in the formation of galaxies and how this compares to that for the major mergers.    Our focus is on the amount of stellar mass which is added to galaxies due to the merger process. 
As such, the way to determine the role of mergers is through measuring the galaxy merger rate.  
 What we do not consider in this paper is the role of minor mergers in producing star formation and AGN activity \citep[][]{Lofthouse2017, Shah2020}. However all of these features depend on how many minor mergers occur within galaxies.

In this subsection, and the next, we discuss both the minor and major merger rates for our sample of galaxies.  We derive these merger rates based on time-scales for mergers to occur which are calculated within a certain stellar mass range, and given the separation of the pair.  This requires a time-scale $\tau_{\text{m}}(z)$ in which the merging system is in a given observed physical state. 

In this section, we describe how we calculate new merger time-scales for pairs in mergers using the Illustris simulation. We then describe in the next section the merger rates for mergers and compare these rates to the merger rates for major mergers which we rederive using these new simulations. 

The important issue with measuring the merger rate is the unknown time-scale for these mergers.  There are various ways in which this can be measured, including empirically \citep[e.g.,][]{Conselice2009a}, or through the use of N-body models which can be used to infer the merger time-scale for different mass ratios, masses and other orbital properties \citep[e.g.,][]{Conselice2006,Lotz2008}.  Values found in these previous studies range from $\tau_{\rm{m}} = 0.3-1$ Gyr for pairs with projected separations of $\leq 30$ kpc.  However, it has also been noted that the merger time-scale evolves with redshift, such that the nominal merger-time declines as $\sim (1+z)^{-2}$, whereby galaxies at higher redshifts merge faster than those at lower redshifts \citep{Snyder2017}.

As this is of major importance for understanding the role of mergers in galaxy assembly, we have taken it upon ourselves to reinvestigate this problem in detail using the IllustrisTNG simulation \citep[][]{Vogelsberger2014}.  We carry this out by selecting all galaxy mergers within the TNG300-1 simulation with stellar mass ratios above $\mu > 0.1$ and with  $M_* > 10^9 M_\odot$ by tracking their relative distances up to the point where their subhalos are considered merged by the sublink algorithm (see \cite{Ferreira2020} for more details). By grouping these mergers in redshift bins based on the snapshot where they merge we fit the merger-time scale $\tau$ as a function of the mass ratio $\mu$, as well as redshift.   

In Figure~4 we show the change in merger time-scales for both major and minor mergers with redshift.  We find that this change is significant over the redshifts we probe in this paper. Ultimately, we find that this change with redshift, and mass ratio of the mergers, is well fit as a power-law of the form:

\begin{equation}
    \tau = a \times (1+z)^{b}
\end{equation}
where the best fit parameters dependence on mass ratio ($\mu$) are:
\begin{equation}\label{eq:timescale_fit}
a = -0.65\pm0.08 \times \mu + 2.06\pm0.01 \\ 
\vspace{0.5cm}
\newline
b = - 1.60\pm0.01
\end{equation}

\noindent as is shown in Figure~\ref{fig:fm-11}.  This goes beyond the calculation of \cite{Snyder2017}, as we have fit how these terms change with both redshift and stellar mass ratio.    We find a less steep relation than \cite{Snyder2017}, such that the slope for equal mass mergers with $\mu = 1$ is $b = -1.19$. However our nominal merger time is lower by about $1.4$ Gyr. Lower mass ratios will have longer merger time-scales, as expected within dynamical friction calculations \citep[e.g.,][]{Conselice2006}.

\begin{figure}
	\includegraphics[width=\columnwidth]{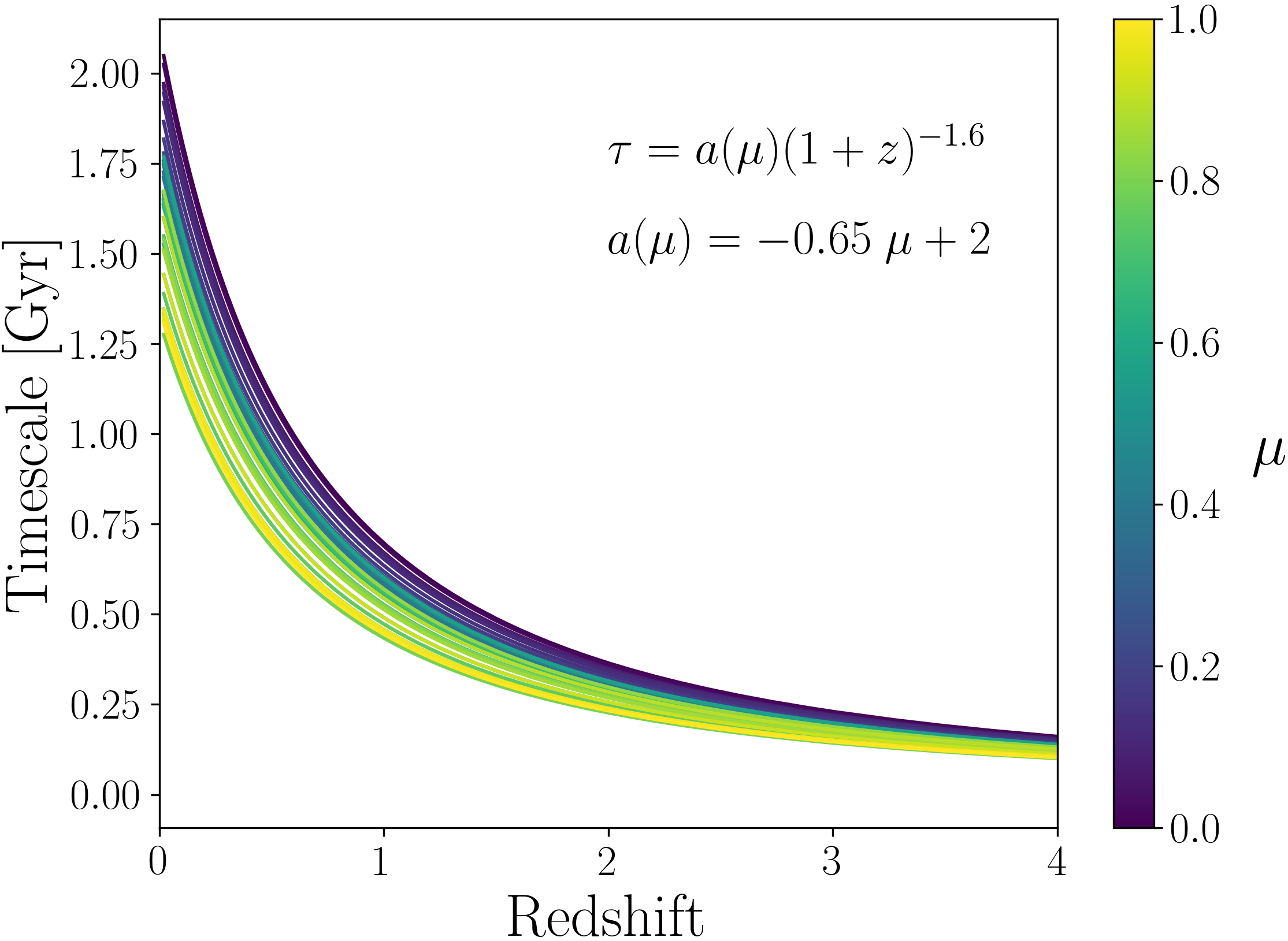}
    \caption{Timescale dependence on redshift and merger mass ratio, $\mu$, based on the best fit parameters $a$ and $b$ of Eq. 27 found by estimating the merging timescale of objects in TNG300-1 with mass ratio $\mu > 0.1$ and with stellar masses $M_* > 10^{9} M_\odot$. }
    \label{fig:fm-11}
\end{figure}

\begin{figure*}
	\includegraphics[width=17cm]{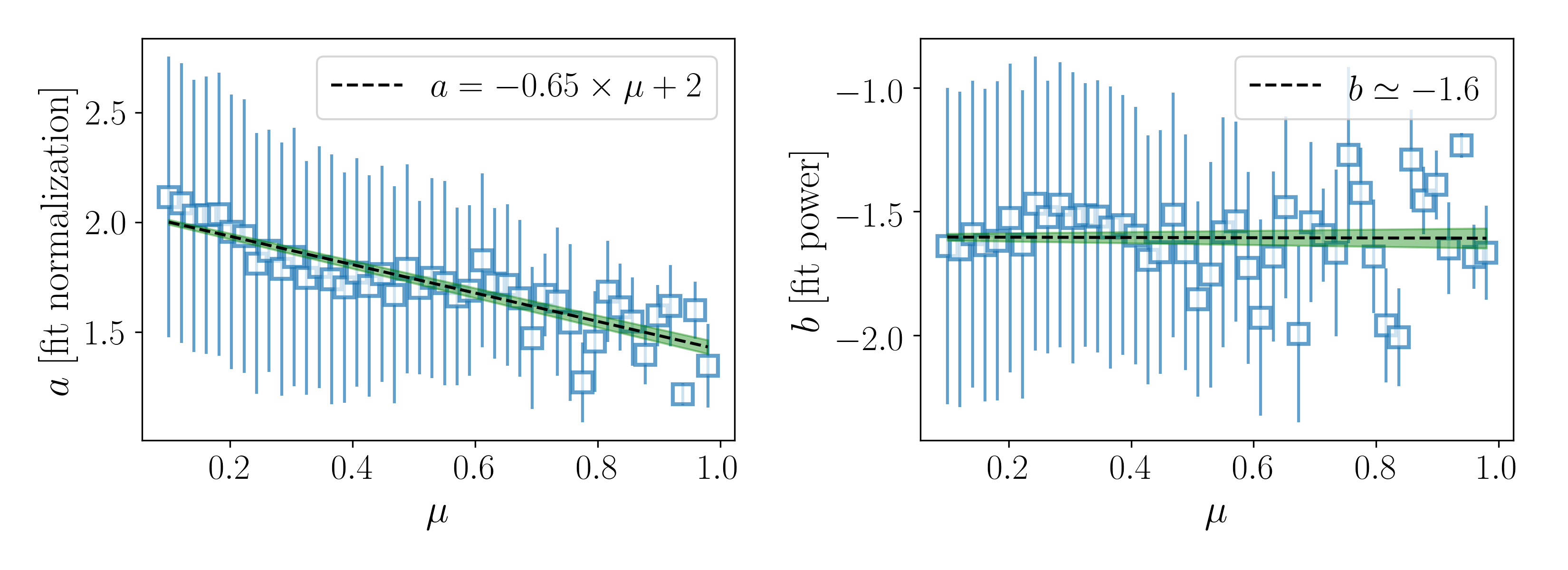}
    \caption{The relation between the fits for the equation for the merger time as a function of the mass ratio $\mu$ given in eq. 8.  The values of $a$ and $b$ are used in Eq. \ref{eq:timescale_fit}, where $a$ shows an inverse dependence on $\mu$, whilst $b$ is found to be independent of $\mu$. Error bars show $\pm 1 \ \sigma$ from the mean values.}
    \label{fig:fm-12}
\end{figure*}

%below is where I have added to the section the previous stuff which was in 4.3

\subsection{Measured Merger Rates}
\label{subsec:merger_rates}

In this section we discuss the galaxy merger rates up to $z \sim 3$, using the time-scales derived in the previous section and the merger fractions from \S \ref{subsec:mergerfraction_evo}.  In a sense, this section is the most important aspect in this paper, as minor mergers have not yet been measured for distant galaxies in the same way that major mergers have been (e.g., M17; \cite[][]{duncan2019}).    This subsection is divided into several parts, all of which display several aspects of the merger history.  

There are a few ways to characterize the merger rates of galaxies.  The two most common are the merger rate per galaxy, $\mathcal{R}$, and the merger rate density $\Gamma$.  The merger rate per galaxy is the most direct one to measure as it involves simply understanding the merger fraction for a sample of galaxies and the time-scale in which that selection is sensitive to the method of detection, which we described in detail in \S \ref{subsec:merger_timescales}.  Otherwise, the galaxy merger rate density is analogous to the star formation rate density  \citep[e.g.,][]{Madau2014} and is essentially the number of galaxy mergers occurring per unit time per unit co-moving volume.  We investigate both of these in the following subsections.

\subsubsection{The Merger Rate per Galaxy}

The formalism we use to calculate the merger rate per galaxy,  $\mathcal{R}$ is very similar to that used in M17 and D19.   In general we calculate the merger rate per galaxy as:

\begin{equation}\label{eq:old_rate_conv}
	\mathcal{R} (>M_*, z) = \frac{f_{P}(>M_*, z) \times C_{\rm merg}}{\tau_{\text{m}}(z)}
\end{equation}

\noindent such that $f_{\text{P}}(>M_*, z)$ is the pair fraction at redshift $z$, with stellar masses greater than a given limit, the value of $C_{\rm merg}$ is fraction of pairs that will actually merge into a single galaxy. Finally the value $\tau_{m} (z)$ the corresponding merger timescale at a given redshift, whose value is calculated from the formulism described in \S \ref{subsec:merger_timescales}.

\begin{figure*}
        \vspace{-1cm}
    \begin{center}
	\includegraphics[width=5in]{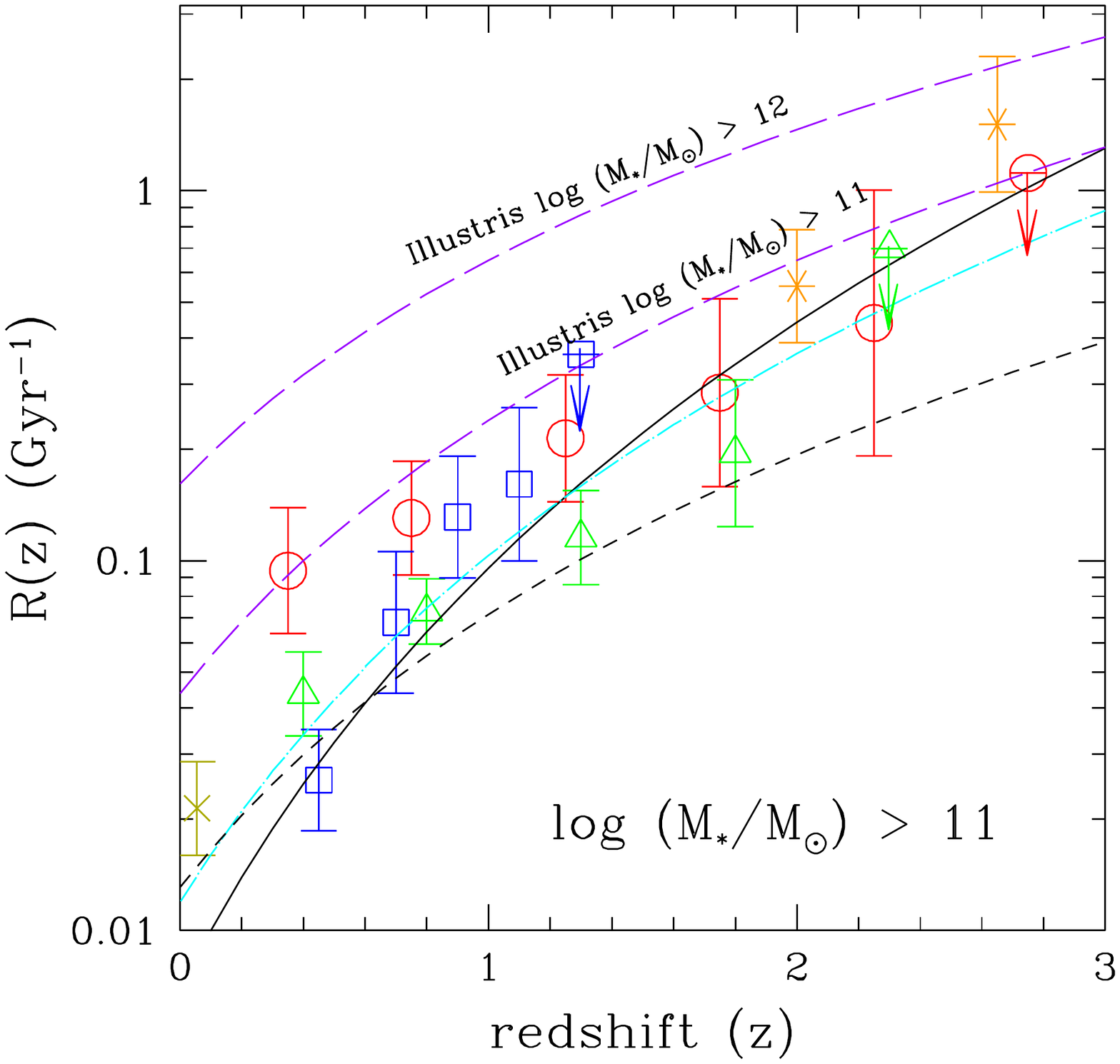}
 	\end{center}
   \label{fig:fm-1em4}
        \vspace{-3cm}
    \caption{Derived total galaxy merger rates, $\mathcal{R}(z)_{\rm tot}$, for galaxies at $M/M_\odot > 10^{11}$ down to stellar mass ratios of 1:10. Estimates are calculated from measured pair fractions in the UDS (red circles), UltraVISTA (green triangles), VIDEO (blue squares) and GAMA (gold crosses) survey fields.  The two orange points at higher redshifts are the merger fractions for minor mergers from Bluck et al. (2012), as in Figure~2. A power-law function (black solid line) is fit to the data points. The dashed line shows the corresponding merger rate for major mergers at the same redshfit range. The minor mergers best fit are shown by the dot-dashed cyan line. }
            \vspace{1cm}

\end{figure*}

We also have to consider the factor, $C_{\rm merg}$, as perhaps not all galaxies that are in pairs will merge within a given time-scale.  The reason for this is that within pairs there are different types of orbits, some of which will result in long merger time-scales.    However, the value will depend to some degree on stellar mass, mass ratio, and redshift.  This value remains uncertain in our merger rate calculations.  For the remainder of this paper we set this value at $C_{\rm merg} = 1$, although we discuss implications for a lower value.  The value for $\tau_{m}$ is calculated with the methodology and models which are discussed in \S \ref{subsec:merger_timescales}.   In the next subsections we describe the merger rate per galaxy at different mass limits for selection.

% The resulting merger rate histories for minor mergers in shown in Figure~5.  We also compare in this plot the merger rates for the minor mergers with those of the major mergers, although we have recalculated the time-scales for this plot from what was published in M17.  

\subsubsection{Merger Rates for Massive galaxies ($>10^{11}\ \mathrm{M}_\odot$)}

Total merger pair fractions for massive ($>10^{11}\ \mathrm{M}_\odot$) galaxies are converted to total merger rates using the methodology explained earlier, inserting our new derivation of the merger time-scale described above. Uncertainties on the merger rates are estimated using a bootstrap approach, incorporating uncertainties on the galaxy stellar mass function (where applicable), the measured pair fractions, cosmic variance estimates, and Poisson noise. 

Estimates of the fractional merger rate, $\mathcal{R}(z)$, are shown in Figure~6.   We find a similar increase with redshift as we do with the merger fraction, although the increase with $(1+z)^{m}$ is steeper due to the fact that the time-scales typically become shorter at higher redshifts, resulting in a higher merger rate than a constant time-scale provides.  A simple power-law best describes the evolution of this rate with redshift for the total merger history, such that: 

\begin{equation}
 \mathcal{R_{\rm tot}}(z, M_{*} > 11) =  (0.7^{+0.3}_{-0.2} \times 10^{-2}) \times (1+z)^{3.77^{+0.34}_{-0.32}} {\rm Gyr^{-1}}
\end{equation}

\noindent giving the best fitting parameters and their uncertainties. This fit is given by the solid curve in  Figure~6. Plotted as the dashed curve is the major merger fractional merger rate. The two curves are similar in shape over redshift and diverge significantly only at high redshift ($z > 2.5$), where the total mergers are higher in value than the major mergers. This suggests that the rate of minor mergers per galaxy per unit time is relatively high compared to major mergers.

We calculate the minor merger rate increase using the same technique used to find the minor merger pair fraction evolution. This is done in the same way by subtracting the major mergers, as previously measured, from the total mergers and then dividing the by the time-scale inferred based on the masses of these pairs.  This then gives us the minor merger rate as, again best fit as a power-law:

\begin{equation}
 \mathcal{R_{\rm minor}}(z, M_{*} > 11) =  (1.2^{+0.4}_{-0.3} \times 10^{-2})  \times (1+z)^{3.10^{+0.38}_{-0.36}} {\rm Gyr^{-1}}
\end{equation}

\noindent For completeness we also discuss the major merger rate, as this will can now be remeasured through our new time-scales.  Thus, the major merger rate for these massive galaxies over time at $z>0$ is:

\begin{equation}
 \mathcal{R_{\rm major}}(z, M_{*} > 11) =  (1.3^{+0.6}_{-0.4} \times 10^{-2})  \times (1+z)^{2.45^{+0.42}_{-0.39}} {\rm Gyr^{-1}}
\end{equation}

\noindent Previous minor merger studies are much less bountiful than those for major mergers. However, estimates of the volume averaged total merger rate including those from \citet[][see their Table 4]{Man2016A} largely agree when we correct for the different time-scales in their work to compare to ours. 

%After their data has been converted to use $C_\text{merg} = 0.6$ and $\left<T_\text{obs}\right> = 0.95$ Gyr (where necessary), their derived merger rates are found to be in agreement with those presented in this work.

Also shown are total merger rate estimates from the Illustris hydrodynamical simulation, given by the dashed purple lines, for stellar masses of M$_{*} > 10^{10}$--$10^{12}\ \mathrm{M}_\odot$.  It is apparent the predictions from the simulation evolve strongly with redshift, increasing from $\sim 5 \times10^{-2}$ Gyr$^{-1}$ at $z = 0$ to $\sim 1$ Gyr$^{-1}$ at $z = 3$. These predictions are therefore qualitatively and quantitatively inconsistent with the derived merger rates of this work at $z > 0.5$. 

%Further estimates of the merger rate from \citet[][see their Table 1]{Man2014} are shown as the black and pink shaded regions in Figure \ref{fig:minor:rates:11}. These represent the $1\sigma$ uncertainty of their fractional merger rates, derived using pair fractions measured with data from the UltraVISTA region, and the 3DHST and CANDELS combination, respectively. This work's merger rates are seen to be in excellent agreement with those of \citeauthor{Man2014}, even out to high redshift.

 \begin{table}
\centering
\caption{Major merger ($\mu > 1/4$) fraction fitting parameters for combinations of survey regions, for a parametrisation of the form $f_\text{pair}(z) = f_0(1+z)^{m}$. Fitting is performed on $f_\text{pair}$ measurements up to the redshifts reported in Tables~2\&3. Errors are determined using a bootstrap analysis and the resulting parameter distributions of 10,000 realisations. The number of merging events, $N_\text{merg}$, a galaxy undergoes at $0 < z < 3.5$, given by the integral in Equation~45, is provided in the far right column.}
\label{tab:fm:fits}

\begin{tabular}{lcccc}
\hline
Type & $f_0$ & $m$ & $N_\text{merg}$ \\
\hline
\multicolumn{4}{c}{$\mathcal{M}_* > 10^{11}\mathrm{M}_\odot\ (5-30\mathrm{kpc})$}\\
\hline
total    & $0.051^{+0.009}_{-0.008}$ & $1.30^{+0.32}_{-0.33}$ & $2.27^{+0.6}_{-0.4}$ \\
major    & $0.024^{+0.005}_{-0.004}$ & $0.85^{+0.19}_{-0.20}$ & $0.84^{+0.3}_{-0.2}$ \\
minor & $0.023^{+0.008}_{-0.007}$ & $1.52^{+0.3}_{-0.33}$ &$1.43^{+0.5}_{-0.3}$ \\
\hline
\multicolumn{4}{c}{$n(>\mathcal{M}_*) = 1 \times 10^{-4}\ $Mpc$^{-3}\ (5-30\mathrm{kpc})$}\\
\hline
total    & $0.042^{+0.011}_{-0.009}$ & $1.09^{+0.34}_{-0.30}$ & $1.38
^{+0.43}_{-0.30}$ \\
major & $0.019^{+0.007}_{-0.006}$ & $1.16^{+0.42}_{-0.37}$ & $0.86^{+0.38}_{-0.25}$ \\
minor & $0.023^{+0.015}_{-0.010}$ & $0.95^{+0.65}_{-0.61}$ & $0.52^{+0.20}_{-0.12}$ \\
\hline
\end{tabular}
\end{table}

%new table may 2022

\begin{table*}
\caption{Fitting parameters for the fractional merger rate, $\mathcal{R}_\text{merg}(z)$ for various combinations of surveys used within this work. Fits with two parameters are of the form $\mathcal{R}_\text{merg}(z) = \mathcal{R}_0 (1+z)^{m_\mathcal{R}}$, while those with three parameters are of the form $\Gamma_\text{merg}(z) = \Gamma_0 (1+z)^{m_\mathcal{R}} \exp(-{c \times z})$. Appropriate fitting forms are decided by comparing the goodness of fit using the $\chi^2$. Parameters and their associated uncertainties are calculating using a bootstrap technique.}
\label{tab:rates:rmerg}
\centering
% UPDATED TO 2017 RESULTS

%the table for the merger rates
\begin{tabular}{lccccc}
\hline
Type & $\mathcal{R}_0$ & $m_\mathcal{R}$ &  $\Gamma_0$ & $m_\mathcal{R}$ & $c_\mathcal{R}$ \\
       & (Gyr$^{-1}$) &  & & (Gyr$^{-1}$ Mpc$^{-3}$)& \\
\hline
\multicolumn{6}{c}{$\mathcal{M}_* > 10^{11}\mathrm{M}_\odot\ (5-30\mathrm{kpc})$} \\
\hline
total & $0.7^{+0.3}_{-0.2} \times 10^{-2}$ & $3.77^{+0.34}_{-0.32}$ & 1.00$\pm0.24 \times 10^{-5}$ & 4.78$\pm$2.09 & -1.95$\pm$0.86 \\
major & $1.3^{+0.6}_{-0.4} \times 10^{-2}$ & $2.45^{+0.42}_{-0.39}$  & 0.36$\pm0.14 \times 10^{-5}$ & 7.39$\pm$0.72 & -3.45$\pm$0.33 \\
minor & $1.2^{+0.4}_{-0.3} \times 10^{-2}$ & $3.10^{+0.38}_{-0.36}$ & 0.45$\pm0.18 \times 10^{-5}$ & 4.41$\pm$1.31 & -1.56$\pm$0.51  \\
\hline

\multicolumn{6}{c}{$n(>\mathcal{M}_*) = 1 \times 10^{-4}\ \mathrm{Mpc}^{-3} (5-30\mathrm{kpc})$} \\
\hline
total & $1.9^{+0.4}_{-0.3} \times 10^{-2}$ & $2.61^{+0.34}_{-0.32}$ & $1.91^{+0.41}_{-0.33} \times 10^{-6}$ & 2.60$^{+0.34}_{-0.32}$ & -\\
major & $1.0^{+0.4}_{-0.4} \times 10^{-2}$ & $2.76^{+0.42}_{-0.39}$ &  $1.03^{+0.39}_{-0.38} \times 10^{-6}$ & $2.76^{+0.41}_{-0.39}$ & -\\
minor & $0.9^{+0.4}_{-0.3} \times 10^{-2}$ & $2.31^{+0.37}_{-0.32}$ & $0.90^{+0.37}_{-0.34} \times 10^{-6}$ & $2.31^{+0.37}_{-0.32}$ & - \\
\hline
\end{tabular}
\end{table*}

\begin{table*}
\centering
\caption{The estimated number density of the primary sample,$n_1$, stellar masses of a typical primary sample galaxy, $\mathcal{M}_{*,1}$, secondary sample galaxy, $\mathcal{M}_{*,2}$, and estimated stellar mass, $\mathcal{M}_*^+$, accreted through major and minor mergers as a function of redshift for two samples of galaxies: $>10^{11}\ \mathrm{M}_\odot$ and $n(>\mathcal{M}_*) = 10^{-4}$ Mpc$^{-3}$.}
\label{tab:arates}
\hspace{-2.5cm} \resizebox{\textwidth}{!}{%
\begin{tabular}{cccccccc}
\hline
\hline
& & & \multicolumn{2}{c}{Major Mergers} & & \multicolumn{2}{c}{Minor Mergers} \\\cline{4-5} \cline{7-8}
$z$ & $n_1$ & $\log \mathcal{M}_{*,1}$ & $\log \mathcal{M}_{*,2}$ & $\log \mathcal{M}_*^+$ & & $\log \mathcal{M}_{*,2}$ & $\log \mathcal{M}_*^+$ \\
 & ($10^{-4}$ Mpc$^{-3}$) & ($\log \mathrm{M}_\odot$) & ($\log \mathrm{M}_\odot$) & ($\log \mathrm{M}_\odot$) & & ($\log \mathrm{M}_\odot$) & ($\log \mathrm{M}_\odot$)  \\
\hline
\multicolumn{8}{c}{$\mathcal{M}_* > 10^{11}\ \mathrm{M}_\odot$} \\
\hline
0.0 -- 0.2 & 3.2$^{+1.3}_{-1.0}$ & 11.2$\pm$0.1 & 10.8$\pm$0.1 & 9.6$\pm$0.1 & & 10.4$\pm$0.1 & 8.8$\pm$0.3 \\
0.2 -- 0.5 & 3.8$^{+1.3}_{-1.0}$ & 11.3$\pm$0.1 & 11.0$\pm$0.1 & 9.9$\pm$0.2 & & 10.5$\pm$0.1 & 9.2$\pm$0.3 \\
0.5 -- 1.0 & 2.7$^{+0.6}_{-0.5}$ & 11.3$\pm$0.1 & 10.9$\pm$0.1 & 10.3$\pm$0.2 & & 10.5$\pm$0.1 & 9.6$\pm$0.3 \\
1.0 -- 1.5 & 1.4$^{+0.3}_{-0.3}$ & 11.2$\pm$0.1 & 10.9$\pm$0.1 &         10.4$\pm$0.2 & & 10.4$\pm$0.1 & 9.8$\pm$0.3 \\
1.5 -- 2.0 & 0.6$^{+0.2}_{-0.2}$ & 11.2$\pm$0.1 & 10.9$\pm$0.1 &         10.5$\pm$0.2 & & 10.4$\pm$0.1 & 9.9$\pm$0.3 \\
2.0 -- 2.5 & 0.3$^{+0.2}_{-0.2}$ & 11.2$\pm$0.1 & 10.8$\pm$0.1 &         10.5$\pm$0.2 & & 10.4$\pm$0.1 & 10.0$\pm$0.4 \\
2.5 -- 3.0 & 0.2$^{+0.2}_{-0.1}$ & 11.2$\pm$0.1 & 10.8$\pm$0.1 & 10.5$\pm$0.2 & & 10.4$\pm$0.1 & 10.1$\pm$0.4 \\
%3.0 -- 3.5 & 0.2$^{+0.6}_{-0.2}$ & 11.2$\pm$0.1 & 10.8$\pm$0.1 & 9.2$\pm$0.2 & & 10.4$\pm$0.1 & 8.8$\pm$0.4 \\ \cline{5-5}\cline{8-8}
           &                     &              &              & 11.2$\pm$0.2 & & & 10.6$\pm$0.3 \\
\hline
\multicolumn{8}{c}{$n(>\mathcal{M}_*) = 1 \times 10^{-4}$ Mpc$^{-3}$} \\
\hline
0.0 -- 0.2 & 1.0$^{+0.6}_{-0.4}$ & 11.3$\pm$0.1 & 10.9$\pm$0.1 & 10.0$^{+0.3}_{-0.2}$ & & 10.5$\pm$0.1 & 9.5$\pm$0.3 \\
0.2 -- 0.5 & 1.0$^{+0.5}_{-0.3}$ & 11.5$\pm$0.1 & 11.2$\pm$0.1 & 10.3$^{+0.3}_{-0.2}$ & & 10.7$\pm$0.1 & 9.7$\pm$0.3 \\
0.5 -- 1.0 & 1.0$^{+0.3}_{-0.2}$ & 11.4$\pm$0.1 & 11.1$\pm$0.1 & 10.5$\pm$0.3         & & 10.6$\pm$0.1 & 9.9$\pm$0.3 \\
1.0 -- 1.5 & 1.0$^{+0.3}_{-0.2}$ & 11.3$\pm$0.1 & 10.9$\pm$0.1 & 10.4$\pm$0.3         & & 10.5$\pm$0.1 & 9.9$\pm$0.4 \\
1.5 -- 2.0 & 1.0$^{+0.3}_{-0.3}$         & 11.1$\pm$0.1 & 10.8$\pm$0.1 & 10.3$\pm0.3$          & & 10.3$\pm$0.1 & 9.8$\pm$0.4 \\
2.0 -- 2.5 & 1.0$^{+0.6}_{-0.5}$ & 11.0$\pm$0.1 & 10.6$\pm$0.1 & 10.1$\pm$0.3         & & 10.2$\pm$0.1 & 9.6$\pm$0.5 \\
2.5 -- 3.0 & 1.0$^{+0.7}_{-0.5}$ & 10.9$\pm$0.1 & 10.5$\pm$0.1 & 9.9$\pm$0.4         & & 10.1$\pm$0.1 & 9.5$\pm$0.5 \\
%3.0 -- 3.5 & 1.0$^{+1.8}_{-0.7}$ & 10.9$\pm$0.1 & 10.6$\pm$0.1 & 9.2$\pm$0.4 & & 10.1$\pm$0.1 & 8.4$\pm$0.6 \\ \cline{5-5}\cline{8-8}
           &                     &              &              & 11.1$\pm$0.3 &      &                & 10.6$\pm$0.4 \\
\hline
\hline
\end{tabular}}%
\end{table*}

\subsection{Volume-averaged merger rate}

%need to write

The volume averaged merger rate is in some ways the ultimate quantity for understanding galaxy mergers. This is particularly true when this quantity is known as a function of redshift (time) as well as stellar mass. This allows us to calculate fundamental quantities such as the number of ongoing mergers, the mass accreted per unit density due to mergers of various types, as well as how this relates as a function of mass. This ultimately will give us a good idea of how mergers are driving the galaxy formation process.  The key aspect here is to use the information on the merger rate per galaxy and to convert this into a merger rate per unit volume.  This is accomplished by examining the number densities of galaxies above a certain mass or number density limit, as this provides the information needed to make this conversion.  The number of mergers per unit volume is also an interesting quantity for understanding the number of gravitational wave sources that we might expect to find in future experiments with e.g., LISA due to the later mergers of the central black holes within these systems.  

We plot the volume averaged merger rate for massive galaxies in Figure~7. We find that the volume-averaged total merger rate for massive galaxies declines $z < 1$, which ranges from $\Gamma \approx 10^{-5} - 10^{-4.3}$ Mpc$^{-3}$ Gyr$^{-1}$ at this redshift regime. At higher redshift, $\Gamma$ is observed to slightly decline out to $z\sim2.25$ beyond which an upper limit constrains the merger rate to $\Gamma < 10^{-4.5}$ Mpc$^{-3}$ Gyr$^{-1}$. The variation in measurements of the merger rates between fields is approximately a factor of $\sim2$ and this ratio remains the same across the redshift range where multiple observations are made. The total mergers are found to be best described by a power-law plus exponential parametrisation, which can be written as:

\begin{eqnarray}\nonumber
%\begin{equation}
\Gamma_{\rm tot}(z, M_{*} > 11) = (1.00\pm0.24 \times 10^{-5}) \times (1+z)^{4.78 \pm 2.09} \\ 
 \times \exp \left( -1.95\pm0.86 \times z \right) {\rm Gyr^{-1} Mpc^{-3}}
\end{eqnarray}
%\end{equation}

\noindent This fit is shown as the solid  curve in Figure~7. This parametrisation better describes the observed evolution of the merger rate compared to a simple power-law used to describe the pair fraction. 

Also plotted is the major ($\mu > 1/4$) merger volume averaged merger rate, given by the dashed curve.  The best fit for this curve is given by:

\begin{eqnarray}\nonumber
%\begin{equation}
\Gamma_{\rm major}(z, M_{*} > 11) =  (0.36\pm0.14 \times 10^{-5}) \\ \times (1+z)^{7.39 \pm 0.72}  
 \times \\ \exp \left( -3.45\pm0.33 \times z \right) {\rm Gyr^{-1} Mpc^{-3}}
\end{eqnarray}
%\end{equation}

As one would expect the total merger rate best-fit line is consistently higher than the major merger rate, however the ratio of major and minor merger rates evolves with redshift. At $z < 0.5$, the total merger rate is a factor of 2--3 larger than the major merger rate, while at $z > 2$, the ratio of total to major merger rate rises steadily to $\sim 2.5$ at $z \sim 3$.

The minor mergers has a similar history and the evolution of $\Gamma$ can be written as:

\begin{eqnarray}\nonumber
%\begin{equation}
\Gamma_{\rm minor}(z, M_{*} > 11) = (0.45\pm0.18 \times 10^{-5}) \\ \times (1+z)^{4.41 \pm 1.31} \times \\
\exp \left( -1.56\pm0.51 \times z \right) {\rm Gyr^{-1} Mpc^{-3}}
\end{eqnarray}
%\end{equation}
All of these fitted forms and shown in Tables~5\,\&\,6 and can be used to calculate various quantities having to do with the assembly of galaxies through merging, as well as the number of events that occur as a function of time and redshift.
\begin{figure*}
        \vspace{-2cm}
    \begin{center}
	\includegraphics[width=5in]{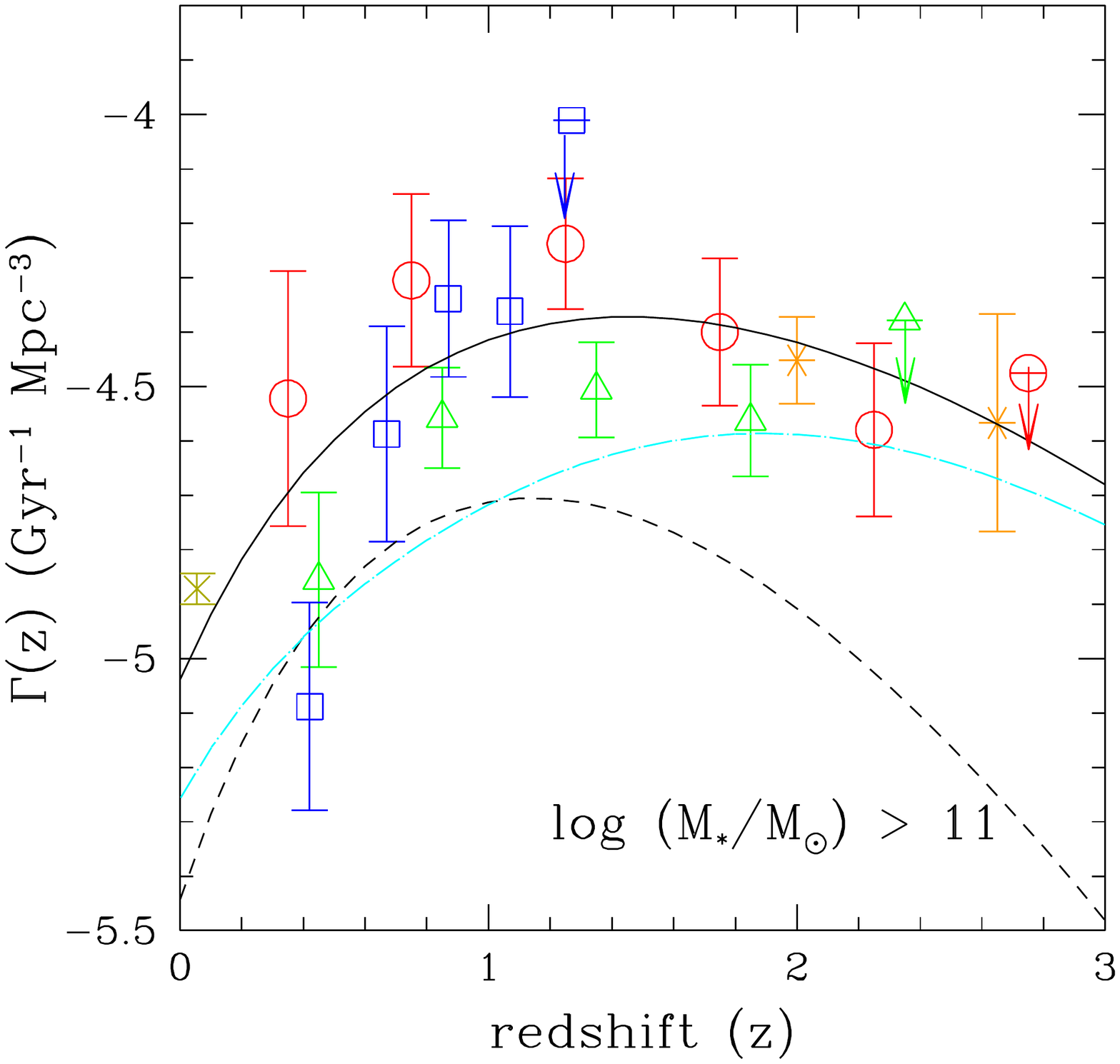}
 	\end{center}
   \label{fig:fm-1em4}
        \vspace{-3cm}
    \caption{Derived volume-averaged merger rate, $\Gamma(z)$ per galaxy, for galaxies at ($M/M_\odot) > 10^{11}$ down to a stellar mass ratio of 1:10. Estimates are calculated from measured pair fractions in the UDS (red circles), UltraVISTA (green triangles), VIDEO (blue squares) and GAMA (gold crosses) survey fields, as in Figure~1.  The orange points at higher redshifts are these measurements from \citep[][]{Bluck2012}. A power-law exponential function (black solid line) is fit to the data points (see text).  The solid black lines show the fit to the overall merger rate per galaxy, while the dashed line shows the major merger rate per galaxy.  The dotted line shows the merger rate per galaxy for the minor mergers. The best fit for the minor mergers merger rate density is shown by the dot-dashed cyan line. }
\end{figure*}

\subsection{Galaxies at a constant number density}

Selecting galaxies at a constant cumulative number density enables the total merger histories of the direct progenitors of local massive galaxies to be probed \citep{Mundy2015}. The larger number density choice explored in M17 is again used to measure the minor merger fractions.    Tracing the progenitors of local massive galaxies with a constant number density selection reveals an equal fraction of galaxies in both major and minor mergers out to $z\sim2.25$, where the fits become unconstrained by observational measurements. This evolution is shown in Figure~8.  Derived values are given by the open symbols for the UKIDSS UDS, VIDEO and COSMOS regions at $z > 0.2$, while gold crosses denote values derived in the GAMA region at low redshift.    

The top left panel of Figure~8 displays the measured total pair fractions for a sample of galaxies selected at a constant cumulative number density of $n(>\mathcal{M}_*) = 10^{-4}$ Mpc$^{-3}$.  The total (solid blue) and minor (cyan dot-dotted) pair fraction is found to increase mildly from a few percent at low redshift to 12\% at $z=2$.  As given by the solid black curve in the figure, this evolution for the constant number density selection ($n$) is described as: 

%june 30, 2021 - edited equation.
\begin{equation}
 f_\text{tot}(z, n) = (0.042^{+0.011}_{-0.009})\times(1+z)^{1.09^{+0.34}_{-0.30}}.
 \end{equation}
 
 \noindent This evolution is generally consistent with that found for galaxies at $>10^{11}\ \mathrm{M}_\odot$, shown as the red dotted line, but with a significantly flatter slope.   Also plotted in the top left of Figure~8 is the major merger pair fraction selected in the same manner, represented by the dashed black curve. As the measured pair fraction slopes are similar, the difference between the major and minor pair fraction for this sample is consistently a factor of $\sim 2$ across most of the redshift range probed.  The minor mergers selected at this constant number density are plotted as the cyan dot-dashed line.  Unlike for the mass selected sample, we find that within this constant number density selection that the rate and fraction of galaxies in major mergers is {\em higher} than that for the minor mergers. 
  
 The top-right panel of Figure~8 displays the evolution of the fractional merger per galaxy rate with redshift (solid line). The total rate is found to increase by a factor of $\sim 15$ from $z = 0$ to $z = 2$, from 0.01 Gyr$^{-1}$ to 0.2Gyr$^{-1}$.  We fit this merger rate history for the constant number density sample as:

% UPDATED TO 2017

\begin{equation}
 \mathcal{R_\text{tot}}(z, n) =  (1.9^{+0.4}_{-0.3} \times 10^{-2}) \times (1+z)^{2.61^{+0.34}_{-0.32}} {\rm Gyr^{-1}}
\end{equation}

\noindent In this same figure, we also show the merger rates for the minor mergers, the major mergers, as well as those selected by a mass limit (red dashed line).   Just as for the merger fractions at a constant number density, we find that the major merger rate selected here is consistently higher than minor merger rate for the same galaxies, which again differs from the mass selected sample.

\begin{figure*}
\vspace{-1.5cm}
	\includegraphics[width=7in]{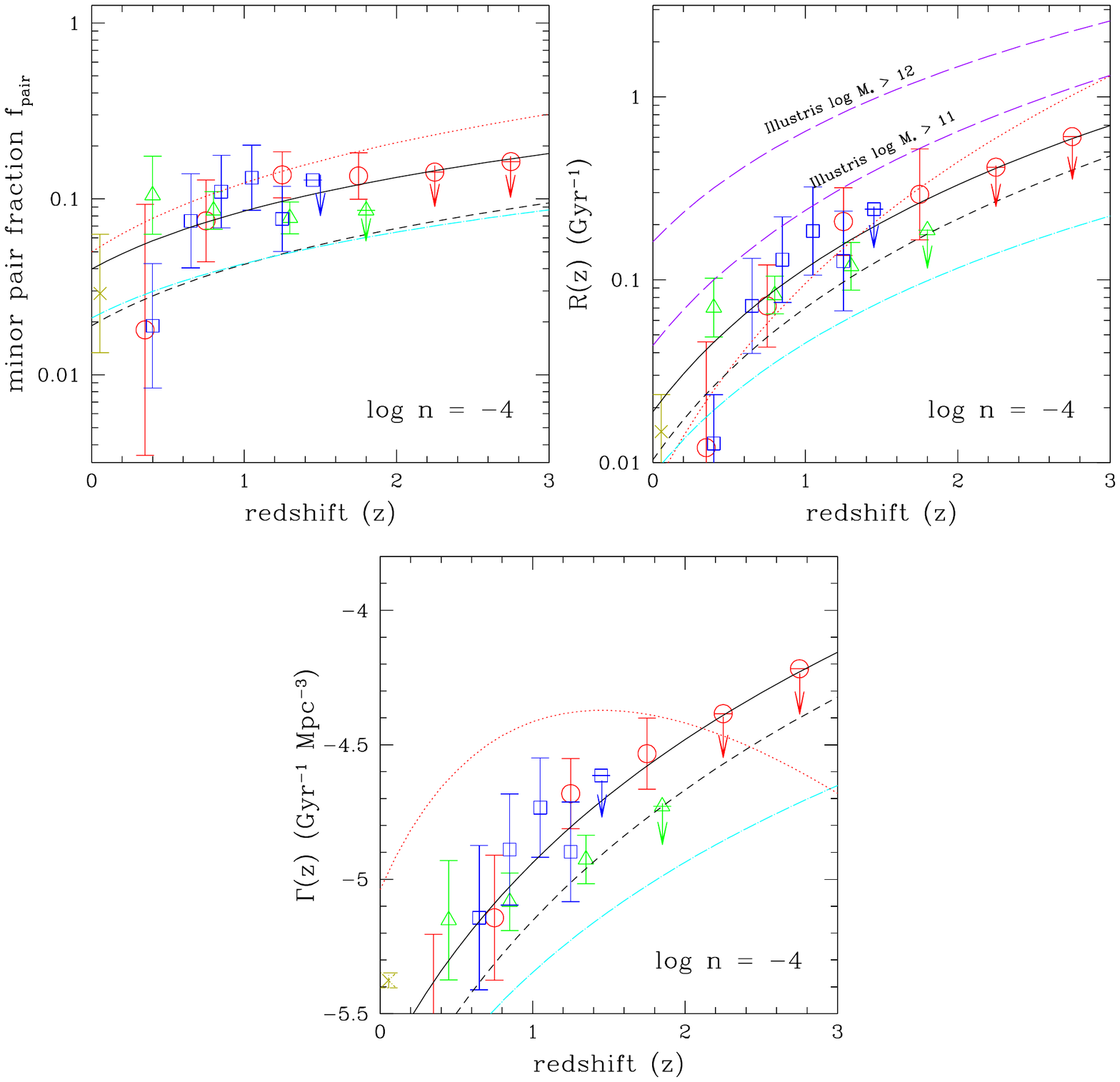}
    \label{fig:fm-1em4}
    \vspace{-2.75in}
    \caption{Constant number density analysis of the merger history for galaxies selected at $n = 1\times10^{-4} \textrm{Mpc}^{-3}$ down to a stellar mass ratio of 1:10.  Show in the three pannels are the measured pair fraction, $f_\text{pair}$ (upper left), galaxy merger rate per galaxy, $\mathcal{R}(z)$ (upper right), and the derived volume-averaged merger rate, $\Gamma(z)$ (bottom). Measurements are made in the UDS (red circles), UltraVISTA (green triangles), VIDEO (blue squares) and GAMA (gold crosses) survey fields. A power law (black solid line) is fit to the total merger history.   The dashed line is the major mergers at the same  $n = 1\times10^{-4} \textrm{Mpc}^{-3}$ selection and the dot-dashed line are for the minor mergers.  The red dotted line is the corresponding massive galaxy selection for systems with M$_{*} > 10^{11}$ \solm.   For the merger rate density plot the power-law exponential function (solid line) is fit to the data points.
}
\vspace{0.5in}
\end{figure*}

The volume-averaged merger rate, displayed in the bottom panel of Figure~8, is observed to increase with redshift from $\sim 10^{-5.5}$ Mpc$^{-3}$ Gyr$^{-1}$ in the lowest redshift bin to $\sim 10^{-4.5}$ Mpc$^{-3}$ Gyr$^{-1}$ in the highest redshift bin which is stellar mass complete. The evolution of this merger rate at constant number density, n, is best parametrised by 

\begin{equation}
 \Gamma_{tot}(z, n) = (1.91^{+0.41}_{-0.33} \times 10^{-6}) \times (1+z)^{2.60^{+0.34}_{-0.32}},
\end{equation}
and is shown in the figure as the solid black curve. Comparing this to the major merger rate, given as the dashed black curve, similar slopes are found but a normalisation approximately a factor of 1.5 larger. This is observed to be constant at redshifts where the fit is constrained by the observational data.   We however find that the minor merger rate density is quite low compared with the major merger rate density. This is due to their being observed few low mass galaxies, as well as the longer time-scale for merging. 

 We investigate the total amount of stellar mass added to galaxies in these different merger selections in quantitative detail in the next section of this paper.  

%this section is ok (Nov 4, 2021)

\subsection{Stellar mass added by mergers}
\label{sfr:mergermass}

Ultimately, this work aims to uncover the role of galaxy mergers in the grander picture of galaxy formation. The stellar mass accrued through major and minor mergers are  important quantities that allows comparisons to be made between other pathways of stellar mass growth, such as star-formation. However, knowing the rate at which a merger event occurs for a given sample of galaxies by itself is not enough to calculate this quantity between two redshifts. The average stellar mass of a companion galaxy must be known as well. 

With this information, the additional stellar mass from mergers, $\mathcal{M}_*^{+}$, for a typical primary sample galaxy between two redshifts can be measured as:

\begin{equation}
\label{eqn:added_mass}
	\mathcal{M}_*^{+} = \int_{t_1}^{t_2} \mathcal{R}_\text{merg}(z)\ \mathcal{M}_{*,2}(z)\ \mathrm{d}t = \int_{t_1}^{t_2} \mathcal{\rho(z)} \mathrm{d}t,
\end{equation}

\noindent where $\mathcal{R}_\text{merg}$ is the fractional merger rate, defined in Equation~29 in terms of the pair fraction, and $\mathcal{M}_{*,2}(z)$ is the average stellar mass of a close-pair companion at redshift $z$. The combination of these two is denoted by $\rho$, with $\rho_{1/4}$ for major mergers, and $\rho_{1/10}$ for minor mergers.

Within any redshift bin the galaxy stellar mass function (GSMF), $\phi(z, \mathcal{M}_*)$, can be used to calculate the average stellar mass of a galaxy within the primary sample, and is defined as
\begin{equation}
\label{eqn:pri_mass}
	\mathcal{M}_{*,1}(z) = \frac{\int^{\mathcal{M}_{*,1}^{\text{max}}}_{\mathcal{M}_{*,1}^{\text{min}}} \phi(z, \mathcal{M}_*)\ \mathcal{M}_*\ \mathrm{d}\mathcal{M}_*}{\int^{\mathcal{M}_{*,1}^{\text{max}}}_{\mathcal{M}_{*,1}^{\text{min}}} \phi(z, \mathcal{M}_*)\ \mathrm{d}\mathcal{M}_*},
\end{equation}

\noindent where $\mathcal{M}_{*,1}^{\text{max}}$ and $\mathcal{M}_{*,1}^{\text{min}}$ are the maximum and minimum stellar masses of the primary galaxy sample, respectively. A similar integration is performed to calculate the average stellar mass of a companion galaxy, $\mathcal{M}_{*,2}(z)$, whereby the integration in Equation \ref{eqn:pri_mass} is instead performed between the stellar mass limits of $\mathcal{M}_{*,1}$ and $\mu \mathcal{M}_{*,1}$.  We calculate the values of these for both minor and major mergers with the integration limits determined by the initial mass of each galaxy (see Table~6).  

The accretion rate history per galaxy differs in substantial ways between the massive galaxy selection and those which are selected at a constant number density, which we discuss in the following sections.

\subsubsection{Major merger stellar mass accretion rate}

Armed with this information we calculate the quantity of stellar mass added through mergers of various types.  Uncertainties are estimated using a bootstrap approach, accounting for errors on the galaxy stellar mass function parameters and the uncertainty in the fit of $f_\text{pair}$.  Note that this calculation is for the amount of stellar mass added due to the merger process from {\em existing} stellar mass. This calculation does not include any star formation that may be induced during the star formation process.   For the major mergers we find the result given on the left-hand panel of Figure~9.  The increase can be parameterized as:

\begin{equation}
\rho_{1/4}(z) = (0.13^{+0.04}_{-0.03} \times 10^{10}) \times (1+z)^{3.05^{+0.40}_{-0.38}}.
\end{equation}

\noindent One of the things we can do with this density change with time is to determine the amount of stellar mass added to galaxies over time due to major mergers. We do this by integrating eq. 40 from $0 < z < 3$.  When we carry out this integration we find that the amount of mass added over this epoch to M$_{*} > 10^{11}$ \solm galaxies is $\delta M_{\rm major,*}$~=~1.48$^{+0.78}_{-0.49}\times 10^{11}$ \solm.  

When we compare with the initial masses of galaxies within this selection, we find that the quantity:

\begin{equation}
    (\delta M_{\rm major,*} / M_{0}) = 0.93^{+0.49}_{-0.31},
\end{equation}

\noindent reveals how much additional mass is added to galaxies over this time period due to major mergers.  Thus, we find that over this epoch that galaxies will almost double in mass during this time, solely due to major mergers.

In Conselice et al. (2022, in prep), we consider the comparison of these stellar mass accretion rates and stellar mass added to galaxies during mergers, to the star formation rate, and the amount of mass added to galaxies in different processes over cosmic time. 

\subsubsection{Minor merger stellar mass accretion rate}

\label{sec:rho_minor_merger}

We use fits to the minor merger pair fraction  and rates discussed earlier in this paper to calculate the minor merger stellar mass accretion rate density. This is done in exactly the same way as described in \S 4.7.1, however the average stellar mass of a companion is calculated over the stellar mass range of $0.1\mathcal{M}_{*,1} < \mathcal{M}_* < 0.25\mathcal{M}_{*,1}$ to obtain the minor merger mass accreted. This represents the number density weighted average stellar mass of a minor merger companion galaxy.  

We fit this increase, plotted in the right panel of Figure~9, in the same way as we have in the previous section for the major mergers. The best fit for this is given by:

\begin{equation}
\rho_{1/10}(z) = (0.022^{+0.010}_{-0.006} \times 10^{10}) \times (1+z)^{3.67^{+0.53}_{-0.46}}.
\end{equation}

\begin{figure*}
	\includegraphics[width=17cm]{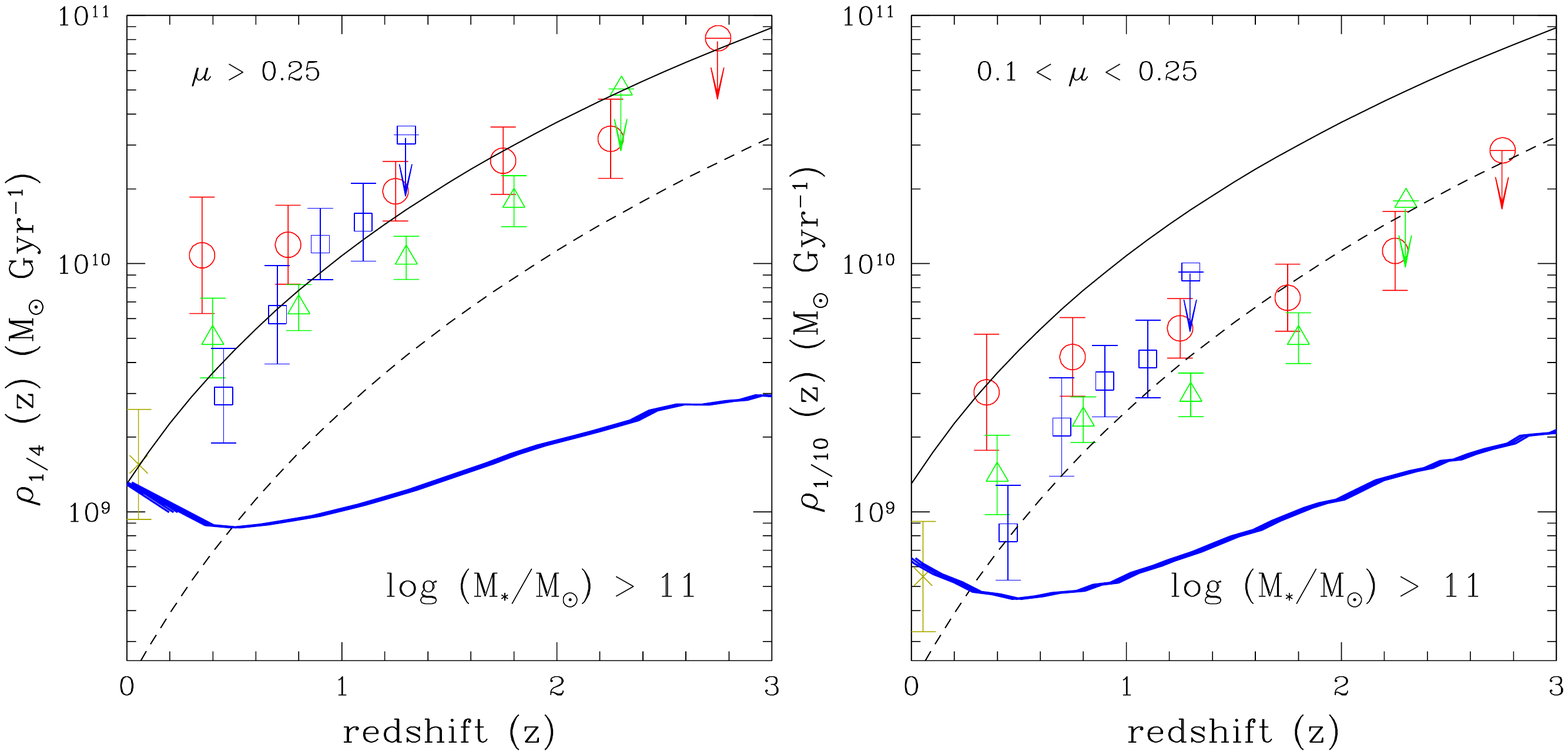} 
	   \vspace{-2.5cm}
    \caption{The evolution of the mass accretion rate due to mergers for M$_{*} > 10^{11}$ \solm selected galaxies.  The left hand side shows the results for the major merger accretion rate due to mergers with mass ratios $\mu > 0.25$, whilst the right side panel shows this relation for the minor mergers with $0.1 < \mu < 0.25$.  The solid line shows the major merger fit and the dashed line is for the minor mergers.  The blue line is the prediction for the accretion rate for mass within systems selected at this mass range from the Illustris simulation (see text). }
    \label{fig:fm-12}
\end{figure*}

\noindent The massive galaxy selection exhibits an increasingly higher minor merger accretion rate density of $10^{9}$ $\mathrm{M}_\odot$ Gyr$^{-1}$ up to several times $10^{10}$ $\mathrm{M}_\odot$ Gyr$^{-1}$. The minor merger mass accretion density is smaller then the major merger rate by up to a factor of a few.  The total amount of mass accreted due to minor mergers is given by:

\begin{equation}
    (\delta M_{\rm minor,*} / M_{0}) = 0.29^{+0.17}_{-0.12},
\end{equation}

This is already a good sign that major mergers are likely the dominate way in which stellar mass is put into galaxies through the merger process, at least for the massive galaxies we study at $z < 3$.

\subsubsection{Constant Number Density}

We also examine the amount of stellar mass accreted due to major and minor mergers for the constant number density selection sample (Figure~10). The approximately constant evolution of $\rho_{1/10}$ and $\rho_{1/4}$ for the constant stellar mass samples is in stark contrast to that seen for the massive selected sample (see Figure~10).  The mass accretion rate  which shows a constant and then perhaps a decline in $\rho_{1/4}$ at high redshift, as opposed to a steadily increasing mass accretion rate due to mergers from the constant stellar mass density selection.  Taking the data points where the pair fraction fits are observationally constrained, $\rho_{1/10}$ is a factor of $\sim5$ smaller than $\rho_{1/4}$ on average. 

In the calculation of this quantity, is it relatively simple  to calculate the expected stellar mass added to a typical galaxy from minor mergers over $z=0-3$. Massive galaxies at a constant number density are found to gain at various redshifts an increase of mass ranging from $\log(M_{*}/\mathrm{M}_\odot) = 9.5 - 9.9$, over $0 < z < 3$.   Below we give the corresponding amount of stellar mass accreted through a constant number density selection.

\begin{equation}
    (\delta M_{\rm major,*} / M_{0}) = 0.11^{+0.13}_{-0.06},
\end{equation}

\begin{equation}
    (\delta M_{\rm minor,*} / M_{0}) = 0.04^{+0.06}_{-0.02},
\end{equation}

\begin{figure*}
	\includegraphics[width=17cm]{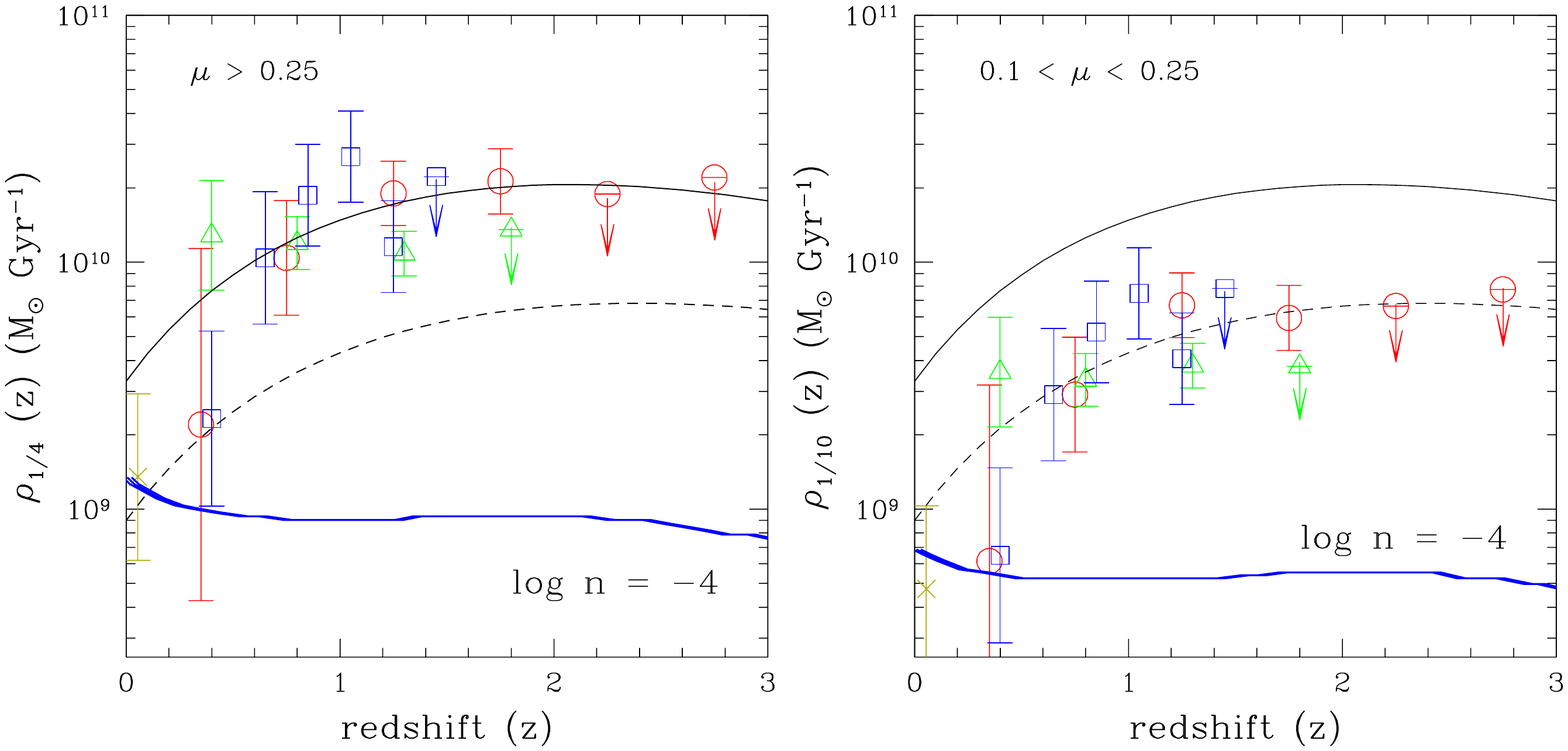}
		   \vspace{-2.5cm}

    \caption{The evolution of the mass accretion rate due to mergers for a constant number density selection at n$>10^{-4}$ Mpc$^{-3}$.  The left hand side shows the data for the major merger accretion rate with mergers with mass ratios $\mu > 0.25$, whilst the right side panel shows this relation for the minor mergers with $0.1 < \mu < 0.25$.  The solid black line shows the major merger fit, and the dashed line is for the minor mergers. The blue line shows the prediction for the accretion rate for mass within systems selected at this number density limit from the Illustris simulation (see text).}
    \label{fig:fm-12}
\end{figure*}

These values agree well with other studies of minor merger histories. Though study of close-pairs, \citet{Man2016A} expect minor mergers to contribute $\log(M_{*}/\mathrm{M}_\odot) = 9.7$ to galaxies with M$_{*} >10^{11}\ \mathrm{M}_\odot$ during $z=0.1-2.5$, with an approximate $\pm80$\% ($\sim0.3$ dex) uncertainty associated with this value. Furthermore \citet{Ferreras2014} study close-pairs of galaxies with $>10^{11}\ \mathrm{M}_\odot$ at $z=0.3-1.3$ in the Survey for High-z Absorption Red and Dead Sources (SHARDS) survey \citep{Perez-Gonzalez2013}. For this selection they find that $(\Delta \mathrm{M} / \mathrm{M}) / \Delta{t} \sim 0.08$ Gyr$^{-1}$, with $\sim$30\% of this mass growth attributed to mergers at mass ratios of 1/4 -- 1/10. For a typical galaxy over this $\sim$5 Gyr period this equates to an upper limit of $\log(\mathrm{M}_\odot) \approx 10.1$ added through `minor' mergers.

\section{Discussion}
\label{sec:discussion}
This paper is an accumulation of various deep imaging and spectroscopic data taken over the past 20 years, and includes results from various studies.  The main aspect is the REFINE survey which is a reanalysis of the UDS, VIDEO, and Ultra-Vista fields.  As such, we have acquired the largest and deepest ground based collection of data in which to examine how the galaxy merger history changes as a function of time.  We have reanalyzed the major mergers, previously studied in M17, as well as providing new analysis of the total mergers, as well as new results on the minor mergers.  As such, we can also determine many new features of galaxy formation, and also make new comparisons with galaxy formation modes.

In general, galaxy formation occurs through a number of ways. The most obvious formation method, and a necessary one, is through star formation driven by converting gas into stars.  We have known however for the last few decades that galaxy formation is not as simple as a conversion of an initial amount of gas into stars over time.  Galaxies start out as typically low mass systems that grow by a rate that is higher than what could be provided by the amount of gas which is in-situ within the galaxies themselves \citep[e.g.,][]{Conselice2012a, Ownsworth2016, padmanabhan2020, walter2020}.  It is thus clear that mergers - both major, minor and ultimately the still unstudied `micro'-mergers of very small galaxies, are an important aspect in the galaxy formation process.  However, mergers and in-situ gas are not enough to sustain the formation we witness, and thus even other processes such as gas accretion are needed to build up the masses of galaxies over time \citep[e.g.,][]{Ownsworth2016, walter2020}.  We can make other conclusions regarding these issues using the results from this paper, which ultimately will feed into an entirely empirical view for how galaxy formation has occurred.  In the following subsections we discuss some of these issues in greater detail.

\subsection{Comparison to Simulations}

There are several features of our results that can be compared with simulations, which we have discussed in passing already.  This includes the merger fraction, merger rates, as well as the amount of stellar mass which is accreted into galaxies as a function of time.  These are important comparisons to make as these models typically are not calibrated on the merger history, but other inferred observational features.  However, within the CDM framework, the merger history of galaxies is a strong prediction, and relies on having a correct understanding of the dark matter temperature, the primordial power spectrum, the dark matter distribution, the baryonic assembly of galaxies and other features.  Thus, this provides a solid test of how we construct these models which are designed to reproduce the galaxy formation process.

We first discuss the agreement, or lack thereof, between the merger fraction measurements and the predictions from various models. As shown in Figures~2 \& 3.  As can be seen by the comparison with the same selection for the H15 models, we find that our measured pair fractions are lower than the simulations for the total mergers.   This differs from the major mergers, discussed in M17, where the merger fractions were more equal.  This is a sign that within this simulation there are more galaxies in pairs with systems which are of lower mass than higher mass systems. This could have implications for predictions of the mass accretion rate, discussed later in this section.  This implies that there are are more close galaxies in this simulation than what is found within the data.  However, what is not clear is how many of these close pairs of galaxies are actually mergers and what types of mergers these are.  This is also different from previous work, whereby the merger fraction predicted in previous incarnations of the Millennium simulation found that the prediction fraction was lower than the merger rates \cite[][]{Bertone2009,Jogee2009}.

Besides the merger or pair fraction, we can also investigate the merger rate - the number of mergers which occur per unit time within these simulations. From Figure~6 and Figure~8 it can be seen that the merger rates within Illustris are actually \textit{higher} than those derived from observational measurements of the merger fraction.  This was also found to be the case in M17, but using a constant time-scale for merging. This suggests that companion galaxies are much less massive than would be expected from an average of the masses of galaxies between the $\mu$ limits (i.e. companion galaxies are at the very low end of the $\mu$ distribution).

Several literature publications are used to calculate the major and minor merger mass accretion rate density, $\rho_{1/4}$, within the Illustris simulation. This is achieved by using fits to the GSMF within the simulation (Equation 1 of \citealt{Torrey2015}) combined with the fitting function of the specific merger accretion rate, $\dot{m}_\text{acc}(\mathcal{M}_*, \mu, z)$, in Table 1 of \citet{Rodriguez-Gomez2016}. It is then in principle straightforward to estimate $\rho$ within Illustris; 

\begin{equation}
\label{eqn:illustris_rho}
\rho^\text{sim} = \int_{\mathcal{M}_{l}}^{\mathcal{M}_{h}} \phi(\mathcal{M}_*, z) \mathcal{M}_*\  \int_{\mu_l}^{\mu_h} \dot{m}_\text{acc}(\mathcal{M}_*, \mu, z)\ \mathrm{d}\mu\ \mathrm{d}\mathcal{M}_*,
\end{equation}

\noindent where $\mathcal{M}_*$ is stellar mass, $\phi(\mathcal{M}_*, z)$ is the GSMF evaluated at $\mathcal{M}_*$ and redshift $z$, and the specific merger accretion rate is defined as

\begin{equation}
\dot{m}_\text{acc}(\mathcal{M}_*, \mu, z) = \frac{1}{\mathcal{M}_*}\frac{\mathrm{d}\mathcal{M}_\text{acc}}{\mathrm{d}t\ \mathrm{d}\mu}.
\end{equation}

\noindent This specific merger accretion rate can then be converted into a mass rate per galaxy (Figure~9 \& 10).  For the purposes of this work an integration is performed with respect to the stellar mass merger ratio between $0.25 < \mu < 1.0$ for major mergers, and $0.1 < \mu < 0.25$ for minor mergers, in order to attain the amount of stellar mass accreted.

Figures 9\,\&\,10 display estimates of the major merger accretion rate density, $\rho_{1/4}$, within the Illustris simulation for galaxies at $>10^{11}\ \mathrm{M}_\odot$ (Figure~9), and galaxies selected at  densities of $n > 10^{-4}$ Mpc$^{-3}$ (Figure~10). The Illustris estimates are given as the lower blue lines in the respective panels. For the constant stellar mass selection (Figure~9), Illustris estimates are typically a factor of $\sim$2-5 times smaller than the estimates based on observations at $z < 2.5$, but generally in agreement at earlier times.  Predictions of $\rho_{1/4}$ for constant number density samples (Figure~10) has a similar shape with the observational estimates, however they are also a factor of 2-5 times smaller over the entire redshift range probed.

%As before, a similarly flat evolution is found for the constant number density sample, remaining at $\approx 5\times10^{-5}$ $\mathrm{M}_\odot$ yr$^{-1}$ Mpc$^{-3}$ across the redshift range probed. 

The differences between the simulation predictions and observational estimates could be due to a number of factors. The ability of simulations to predict the observed number densities of massive galaxies will affect not only the primary sample, but the average stellar mass of an accreted secondary sample companion.  Furthermore, M17 also found that while the merger rates in Illustris are over predicted, the mass accretion rates were underpredicted, as we find here.   At $z < 4$ it has been shown that Illustris matches observations of the GSMF at masses of $10 < \log(\mathcal{M}_*/\mathrm{M}_\odot) < 11$ however it typically \textit{over}-predicts the number densities of galaxies with $<10^{10}\ \mathrm{M}_\odot$ by 0.3--0.5 dex on average, increasingly at lower masses \citep[][see their Fig.~3]{Genel2014}.  These low mass galaxies would dominate the sample, and lower the level of the amount of accreted mass compare to the observations.  There is also evidence that at our redshifts of interest there are fewer massive galaxies, which means that the massive systems selected are lower in mass than the observed systems.    This is particularly a problem with the Illustris simulation we use which has a smaller volume than our observations - 100 cMpc - which limits the number of very massive galaxies, and thus a lower mass accretion rate for these galaxies is obtained as the amount of mass increase is modest compared to our results. When these slightly lower mass galaxies accrete systems of even lower mass, the resulting predicted mass accretion rate would be lower than what we observe.  It is interesting that the rate of accretion at low redshift, and in particular at $z =0$ matches well the predictions, but deviates increasingly at higher redshifts. There may be an issue of mismatched time-scales within these comparisons as well.

\subsection{Number of merger events at $0< z < 3.5$}

The total number of merging events from major and minor mergers are measured using the same procedure as in M17.   This quantity, $N_\text{merg}$, can be calculated for different redshift ranges and survey region combinations. From these quantities it is trivial to estimate the contributions from major and minor mergers towards the total number of merging events over the past 11 Gyr.  We however only give a summary of the number of major and minor merger events which have occurred over the epoch $0 < z < 3$.  We integrate the merger rate to obtain the number of mergers, $N_\text{merg}$, between two redshift bins, which is given by:

\begin{equation}
	\label{eqn:n_merg}
	N_\text{merg} = \int_{t_1}^{t_2} \mathcal{R}_\text{merg}(z)\ dt = \int_{z_1}^{z_2} \frac{\mathcal{R}_\text{merg}(z)}{(1+z)H(z)}\ \mathrm{d}z,
\end{equation}
where we use $\mathrm{d}t = \mathrm{d}z / (1+z)H(z)$, and $H(z)$ is the Hubble constant at redshift $z$, which can be defined as defined as $H(z) = H_0(\Omega_M(1+z)^3 + \Omega_\Lambda)^{1/2}$.   Where the results of these integrals are shown in Figure~4. 

Comparing the calculated total number of merger events at $z < 3$ to the major merger events presented in \S \ref{subsec:merger_rates}, we find that a typical galaxy at $>10^{11}\ \mathrm{M}_\odot$ experiences 0.84$^{+0.3}_{-0.2}$ major mergers and 1.43$^{+0.5}_{-0.3}$ minor mergers over this time. Within this population, on average almost every galaxy experiences a major merger, and on average every galaxy experiences a minor merger.   Finally, the constant number density sample of galaxies undergo 0.5$^{+0.4}_{-0.2}$ major mergers and 0.2$^{+0.3}_{-0.1}$ minor mergers per galaxy at $z < 3$.

There are only a few previous works where we can make a direction comparison to these results. For example, \citep[][]{Man2016A} calculate the expected number of minor mergers for galaxies at M$_{*} >10^{10.8}\ \mathrm{M}_\odot$ selected by stellar mass ratio. Converting the estimated number of mergers (see their Table 5)  \citep[][]{Man2016A} find massive galaxies undergo 0.7-1.1 minor mergers at $0.1 < z < 2.5$.  This is within the errors the same rate that we have calculated, showing a consistency between different studies of this process.

\section{Conclusions}
\label{sec:conclusion}
We present the results of determining how mergers of various types are driving the stellar mass assembly of galaxies within the universe at $0< z < 3$.  It has long been speculated that the mergers  of galaxies, whereby two existing galaxies which have their own previous star formation and merger history, merge themselves into a new system is a significant part of this galaxy formation process.  Mergers are in fact perhaps a dominant method by which galaxies form and grow throughout the history of the universe.  Although there have been a considerable number of studies on this process, we are still learning the basics of this problem and what the answers to ultimate questions concerning galaxy formation might be. 

Major mergers have been the primary method for measuring the merger history  \citep[e.g., M17,][]{duncan2019,Whitney2021}, yet this is only one way in which galaxies can assemble through the merger process.  Mergers of galaxies with other galaxies of significantly lower mass (25\% or less of the primary) with themselves are thought to be a major way in which galaxy formation occurs through cosmic history. However, these types of ``minor mergers'' have not been measured in any detail throughout the history of the universe, thus it is an important missing piece of the observational puzzle for how galaxy formation occurs.  

To address this question we probe in this paper the minor merger ($1/10 < \mu < 1/4$) histories of massive galaxies by measuring the total merger ($\mu > 1/10$) pair fractions of massive galaxies at $z < 3$ using the same technique presented in M17 for major mergers.  We use new derivations of merger time scales and further compare a revised estimate of the major and total merger history for galaxies at $z < 3$.  The minor merger histories of galaxies at $>10^{11}\ \mathrm{M}_\odot$ and at a constant cumulative number density of $n(>\mathcal{M}_*) = 10^{-4}$ Mpc$^{-3}$ have been measured for the first time over a large area of the sky.  \\

\noindent We summarise our findings as follows: \\

\noindent I. We measure the total merger pair fractions at $z < 0.2$ in the GAMA region, and at $0.2 < z < 3$ using a combination of the UKIDSS UDS, VIDEO, and Ultra-VISTA survey regions as part of the REFINE survey. This provides a probed area of 144 sq. deg. at $z < 0.2$, and an area of 3.25 square degrees at $0.2 < z < 3.5$ in which we probe the pair fractions of massive galaxies. 

\noindent II. We find that the measured total minor pair fractions for galaxies at $>10^{11}\ \mathrm{M}_\odot$ are roughly twice as high as that observed for major mergers across the entire redshift range we probe. Furthermore, an approximately constant pair fraction of $\sim20\%$ predicted by the \citet{Henriques2014} semi-analytical model is not consistent with observed pair fractions at $z < 1.5$, and is not qualitatively consistent with the observed slope found for our sample. Similar differences are found with the intermediate stellar mass sample ($>10^{10}\ \mathrm{M}_\odot$) which exhibit total minor pair fractions a factor of 2--4 times larger than the major merger pair fractions. However, no comparisons can be made at redshifts higher than $z = 1.5$ at these lower masses until deeper observations are made. 

\noindent III. We derive the total merger rates per galaxy for our massive galaxy selected sample. The derived total merger rates for massive galaxies imply that minor mergers occur about two times more often as major mergers over the evolutionary history of a typical galaxy over the last 11 Gyr.  Quantitatively, over the redshift range from $0 < z < 3$ there are 0.84$^{+0.3}_{-0.2}$ major mergers and 1.43$^{+0.5}_{-0.3}$ minor mergers for massive galaxies.  

\noindent IV. We furthermore calculate the number density of mergers, or the merger rate density. We find that this quantity increases at higher redshifts but turns over at the highest redshifts.  As the galaxy merger rate per galaxy continues to increase, this decline is simply due to there being fewer massive galaxies at high redshift.  For the number density selected sample, we see a continual increase in the galaxy merger rate density up to the highest redshifts where it can be measured. 

\noindent V. We calculate the mass accretion rate due to mergers due to both minor and major as a function of redshift over $0 < z < 3$.  We find a steadily increasingly mass accretion rate for both minor and major mergers for the massive galaxy selection, but a more gradual increase over the same time-period for the constant number density selected systems. Overall, the total net effect of this increase is to increase the stellar masses of these galaxies by $\sim 120$\%.  Comparing to models we find that the accretion rate of mass is about fives times lower in the models than what we observe.  
 
It is clear from the results presented in this paper that high-redshift measurements are needed to constrain the minor merger histories of massive galaxies at these regimes. This requirement was met in part by the results of Duncan et al. (2019) who probe mergers at a range of stellar mass ratios at $2 < z < 6$ using the full set of CANDELS fields. However, larger fields are required to minimise interference from the observed cosmic variance identified in this work, as well as others \citep[e.g.,][]{Man2014}. 

This work show the importance of mergers in the galaxy formation process.  Because mergers induce central black hole grown and AGN activity, as well as  star formation, the observational history of mergers is an essential first step towards understanding these other fundamental processes and how they occur.  Another major conclusion from this paper is that galaxy formation can be studied observationally -- not just through the comparison to models -- but directly through empirical observations of the formation process. 

Furthermore, galaxy formation models are typically not designed to match observations of merger histories, thus these comparisons are a new and powerful test of galaxy formation models and ideas.  From these observations, and future ones of mergers with telescopes such as Euclid, Rubin/LSST, and JWST we will be able to perform similar experiments to this, but with much better observations over a factor of five thousand times larger area with Euclid, and to a much deeper depth with JWST.  With this data we will answer with a high accuracy the merger history and thus the empirical galaxy formation history of galaxies at all epochs. 

\section*{Acknowledgements}

We thank the entire teams that produced the public data we utilize in this paper, including the Ultra-VISTA, VIDEO, UDS, and GAMA surveys.  Support from the University of Nottingham, the University of Manchester, and the Leverhulme Trust is gratefully acknowledged.  Some of this work was supported by the ERC Advanced Investigator Grant EPOCHS (788113) from the European Research Council (ERC) (PI Conselice), and by STFC through studentships to CJM and KJD. KJD acknowledges funding from the European Union’s Horizon 2020 research and innovation programme under the Marie Sk\l{}odowska-Curie grant agreement No. 892117 (HIZRAD).   For the purpose of open access, the author has applied a Creative Commons Attribution (CC BY) licence to any Author Accepted Manuscript version arising from this submission.

%% An example figure call using \includegraphics

%\acknowledgments

%Acknowledge people, facilities, and software here but remember that this counts
%against your 1000 word limit.

\bibliography{corrected_refs}{}
\bibliographystyle{aasjournal}



\end{document}